\documentclass[12pt]{article}
\pdfoutput=1
\usepackage{epsfig,amsfonts,amsthm}
\usepackage[normalem]{ulem}
\usepackage{amsmath,amssymb}
\usepackage{array}
\usepackage{amsmath}
\usepackage{amsfonts}
\usepackage{amssymb}
\usepackage{subfig}
\usepackage{wrapfig}
\usepackage{graphicx}
\newcommand{\be}{\begin{equation}}
\newcommand{\ee}{\end{equation}}
\newcommand{\bea}{\begin{eqnarray}}
\newcommand{\eea}{\end{eqnarray}}

\renewcommand{\Re}{\mathrm{Re }}
\renewcommand{\Im}{\mathrm{Im }}
\newcommand{\doublet}[2]{ \left( \begin{array}{c}#1 \\ #2 \end{array}\right) }

\usepackage{color}

\def\lsim{\mathrel{\rlap{\lower4pt\hbox{\hskip1pt$\sim$}}
    \raise1pt\hbox{$<$}}}         
\def\gsim{\mathrel{\rlap{\lower4pt\hbox{\hskip1pt$\sim$}}
    \raise1pt\hbox{$>$}}}         

\def\beq{\begin{equation}}
\def\eeq{\end{equation}}
\def\bea{\begin{eqnarray}}
\def\eea{\end{eqnarray}}

\def\<{\left\langle}
\def\>{\right\rangle}

\newcommand{\GeV}{{\ensuremath\rm \,GeV}}

\usepackage[normalem]{ulem}

\def\lsim{\mathrel{\rlap{\lower4pt\hbox{\hskip1pt$\sim$}}
    \raise1pt\hbox{$<$}}}         
\def\gsim{\mathrel{\rlap{\lower4pt\hbox{\hskip1pt$\sim$}}
    \raise1pt\hbox{$>$}}}         

\def\beq{\begin{equation}}
\def\eeq{\end{equation}}
\def\bea{\begin{eqnarray}}
\def\eea{\end{eqnarray}}

\def\<{\left\langle}
\def\>{\right\rangle}

\newcommand{\gev}{\mathrm{\;GeV}} 
\newcommand{\bt}{\begin{tabular}}
\newcommand{\et}{\end{tabular}}

\usepackage{hyperref}

\allowdisplaybreaks[2]
\begin{document}
\bibliographystyle{OurBibTeX}

\title{\hfill ~\\[-30mm]
                  \textbf{CP violating scalar Dark Matter
                }        }
\date{}

\author{\\[-5mm]
A. Cordero-Cid\footnote{E-mail: {\tt adriana.cordero@correo.buap.mx}} $^{1}$,\ 
J. Hern\'andez-S\'anchez\footnote{E-mail: {\tt jaime.hernandez@correo.buap.mx}} $^{1}$,\ 
V. ~Keus\footnote{E-mail: {\tt Venus.Keus@helsinki.fi}} $^{2,3}$,\ 
S. ~F.~King\footnote{E-mail: {\tt King@soton.ac.uk}} $^{3}$,\ \\
S. ~Moretti\footnote{E-mail: {\tt S.Moretti@soton.ac.uk}} $^{~3,4}$,\
D. ~Rojas\footnote{E-mail: {\tt drojas@ifuap.buap.mx}} $^{1}$,\  
D. ~Soko\l{}owska \footnote{E-mail: {\tt Dorota.Sokolowska@fuw.edu.pl}} $^{~~5}$
\\ \\
\emph{\small $^1$ Facultad de Ciencias de la Electr\'onica, Benem\'erita Universidad Aut\'onoma de Puebla,
}\\ 
 \emph{\small  Apdo. Postal 542, C.P. 72570 Puebla, M\'exico,}\\
  \emph{\small $^2$ Department of Physics and Helsinki Institute of Physics,}\\
 \emph{\small Gustaf Hallstromin katu 2, FIN-00014 University of Helsinki, Finland}\\
  \emph{\small $^3$ School of Physics and Astronomy, University of Southampton,}\\
  \emph{\small Southampton, SO17 1BJ, United Kingdom}\\
  \emph{\small  $^4$ Particle Physics Department, Rutherford Appleton Laboratory,}\\
 \emph{\small Chilton, Didcot, Oxon OX11 0QX, United Kingdom}\\
  \emph{\small  $^5$ University of Warsaw, Faculty of Physics, Pasteura 5, 02-093 Warsaw, Poland.}\\[4mm]
  }

\maketitle

\vspace*{-10mm}

\begin{abstract}
\noindent
{We study an extension of the Standard Model (SM) in which two copies of the SM scalar $SU(2)$ doublet which do not acquire a Vacuum Expectation Value (VEV), and hence are \textit{inert}, are added to the scalar sector. We allow for CP-violation in the \textit{inert} sector, where the lightest \textit{inert} state
is protected from decaying to SM particles through the conservation of a $Z_2$ symmetry. The lightest neutral particle from the \textit{inert} sector, which has a mixed CP-charge due to CP-violation, is hence a Dark Matter (DM) candidate. We discuss the new regions of DM relic density opened up by CP-violation,
and compare our results to the CP-conserving limit and the Inert Doublet Model (IDM). 
We constrain the parameter space of the CP-violating model using recent results from the Large Hadron Collider (LHC) and DM direct and indirect detection experiments. } 
 \end{abstract}
\thispagestyle{empty}
\vfill
\newpage
\setcounter{page}{1}

\section{Introduction}
In 2012 both ATLAS and CMS experiments at the CERN Large Hadron Collider (LHC) reported \cite{Aad:2012tfa,Chatrchyan:2012ufa} the observation of a scalar boson with a mass of $\approx 125$ GeV. Although the properties of the observed boson are in accordance with those of the Higgs boson of the Standard Model (SM), it remains an intriguing possibility that it may just be one member of an extended scalar sector. Even though so far no signs of detection of physics Beyond SM (BSM) have been reported, it is well understood that the SM of particle physics is incomplete. A good motivation for BSM is the lack of a Cold Dark Matter (CDM) candidate in the SM.

Although the nature of Dark Matter (DM) is not yet known, according to the 
Standard Cosmological $\Lambda$-CDM Model \cite{Ade:2015xua} it should be a
particle which is stable on cosmological time scales, cold (i.e., non-relativistic at the onset of galaxy formation), non-baryonic,  neutral and weakly interacting. Various such candidates for a state with these characteristics exist in the literature, the most well-studied being the Weakly Interacting Massive Particles (WIMPs) \cite{Jungman:1995df,Bertone:2004pz,Bergstrom:2000pn}, with masses between a few GeV and a few TeV. Any such WIMP candidate must be cosmologically 
stable, usually due to the conservation of a discrete symmetry,
and must freeze-out (i.e., drop out of thermal equilibrium) to result in the 
observed relic density \cite{Ade:2015xua}
:
\begin{equation}\label{relic}
\Omega_{\rm DM} h^2 = 0.1199 \pm 0.0027. 
\end{equation}

It is clear that the SM scalar sector cannot provide a WIMP candidate. However, it was suggested some time ago that the scalar sector could be extended by the addition of an extra doublet, which may not develop
a Vacuum Expectation Value (VEV) while leaving a discrete $Z_2$ symmetry unbroken \cite{Deshpande:1977rw}.
This possibility, which is known as the Inert Doublet Model (IDM),
has been studied extensively for the last few years (see, e.g., \cite{Ma:2006km,Barbieri:2006dq,LopezHonorez:2006gr}). 
Since the IDM involves {\em 1} Inert Doublet plus {\em 1} active Higgs Doublet, we shall also refer to it henceforth as the I(1+1)HDM. 

In the IDM, aka the I(1+1)HDM, one extra spin-zero $SU(2)_L$ doublet with the same SM quantum numbers as the SM-Higgs doublet is added to the scalar sector. One of the possible vacuum states in this model involves the first doublet acquiring a VEV is referred to as the \textit{active doublet}, while the second doublet does not develop a VEV and is henceforth called the \textit{inert doublet} since it does not take part in Electro-Weak Symmetry Breaking (EWSB). This doublet does not couple to fermions and it is by construction the only $Z_2$-odd field in the model, therefore, it provides a stable DM candidate, namely the lightest state among scalar and pseudo-scalar $Z_2$-odd particles.

The I(1+1)HDM remains a viable model for a scalar DM candidate, being in agreement with current experimental constraints. As of now, there are two regions of DM masses where one can expect viable solutions: a low DM mass region, $53 \gev \lesssim m_{\rm DM} \lesssim m_W$ and a heavy DM mass region, $m_{\rm DM} \gtrsim 525 \gev$. The most recent experimental data, both from direct detection experiments and from the LHC, has reduced the viable parameter space in the low mass region \cite{Krawczyk:2013jta,Arhrib:2013ela,Ilnicka:2015jba}. However, in the heavy mass region where the sensitivity of DM direct detection experiments decreases significantly with increasing DM mass, the DM candidate may escape possible detection in the I(1+1)HDM.

In recent papers \cite{Keus:2014jha,Keus:2015xya} we studied DM in a CP-conserving model with {\em 2} inert Higgs plus {\em 1} active Higgs doublet, which we referred to as the I(2+1)HDM. 
We showed that in the light mass region $(m_{\rm DM} \lesssim m_W)$ the extended scalar sector can relax the exclusion limits from direct detection experiments, providing a viable DM candidate in a region of parameter space which would be excluded in the I(1+1)HDM. 
In the heavy DM mass region, we showed that heavy Higgs DM becomes more readily observable as a result of either lowering the DM mass to $360 \gev \lesssim m_{\rm DM}$, or increasing the DM-Higgs coupling, or both,
while always maintaining the DM relic density within the required region. 

In the present paper we look into the CP-violating I(2+1)HDM. CP-violation is introduced in the \textit{inert} sector. Note that the \textit{inert} sector is protected by a conserved $Z_2$ symmetry from coupling to the SM particles, therefore, the amount of CP-violation introduced here is not constrained by SM data. The third and \textit{active} doublet in our model has exactly the same couplings as the SM-Higgs doublet hence the CP-violation in the \textit{inert} sector does not affect the SM-Higgs couplings\footnote{Introducing CP-violation into the \textit{active} sector is restricted by many SM data, for a relevant recent paper, for example, see \cite{Keus:2015hva}.}.

The layout of the remainder of this paper is as follows. In Section \ref{scalar-potential} we present the scalar potential and the mass spectrum. In Section \ref{constraints} we impose all theoretical and experimental constraints on the parameter space of the model. In Section \ref{coannihilation} we introduce the benchmark scenarios relevant for DM studies. In Section \ref{numerical-analysis} we present our numerical analysis for chosen benchmark scenarios and in Section \ref{conclusion} we draw our conclusions.

\section{The scalar potential}
\label{scalar-potential}

It has been shown in \cite{Ivanov:2011ae} that an 3-Higgs-Doublet Model (3HDM) potential symmetric under a group $G$ of phase rotations can be divided into two parts;
a phase invariant part, $V_0$, and a collection of extra terms ensuring the symmetry group $G$, $V_G$.

We now construct our $Z_2$-symmetric 3-Higgs Doublet Model potential, under which the three Higgs doublets $\phi_{1,2,3}$ transform, respectively,  as: 
\be 
\label{generator}
g_{Z_2}=  \mathrm{diag}\left(-1, -1, 1 \right). 
\ee
The resulting potential is of the following form\footnote{Note that adding extra $Z_2$-respecting terms such as
$ 
(\phi_3^\dagger\phi_1)(\phi_2^\dagger\phi_3), 
(\phi_1^\dagger\phi_2)(\phi_3^\dagger\phi_3), 
(\phi_1^\dagger\phi_2)(\phi_1^\dagger\phi_1)$ and/or 
$(\phi_1^\dagger\phi_2)(\phi_2^\dagger\phi_2)
$
does not change the phenomenology of the model. The coefficients of these terms, therefore, have been set to zero for simplicity.}:
\bea
\label{V0-3HDM}
V_{3HDM}&=&V_0+V_{Z_2}, \\
V_0 &=& - \mu^2_{1} (\phi_1^\dagger \phi_1) -\mu^2_2 (\phi_2^\dagger \phi_2) - \mu^2_3(\phi_3^\dagger \phi_3) \nonumber\\
&&+ \lambda_{11} (\phi_1^\dagger \phi_1)^2+ \lambda_{22} (\phi_2^\dagger \phi_2)^2  + \lambda_{33} (\phi_3^\dagger \phi_3)^2 \nonumber\\
&& + \lambda_{12}  (\phi_1^\dagger \phi_1)(\phi_2^\dagger \phi_2)
 + \lambda_{23}  (\phi_2^\dagger \phi_2)(\phi_3^\dagger \phi_3) + \lambda_{31} (\phi_3^\dagger \phi_3)(\phi_1^\dagger \phi_1) \nonumber\\
&& + \lambda'_{12} (\phi_1^\dagger \phi_2)(\phi_2^\dagger \phi_1) 
 + \lambda'_{23} (\phi_2^\dagger \phi_3)(\phi_3^\dagger \phi_2) + \lambda'_{31} (\phi_3^\dagger \phi_1)(\phi_1^\dagger \phi_3),  \nonumber\\
 V_{Z_2} &=& -\mu^2_{12}(\phi_1^\dagger\phi_2)+  \lambda_{1}(\phi_1^\dagger\phi_2)^2 + \lambda_2(\phi_2^\dagger\phi_3)^2 + \lambda_3(\phi_3^\dagger\phi_1)^2  + h.c. \nonumber
\eea
The parameters of the $V_0$ part of the potential are by construction real. We allow for the parameters of $V_{Z_2}$ to be complex, hence introducing explicit CP-violation in the model.

The doublets are defined as
\be 
\phi_1= \doublet{$\begin{scriptsize}$ H^+_1 $\end{scriptsize}$}{\frac{H^0_1+iA^0_1}{\sqrt{2}}},\quad 
\phi_2= \doublet{$\begin{scriptsize}$ H^+_2 $\end{scriptsize}$}{\frac{H^0_2+iA^0_2}{\sqrt{2}}}, \quad 
\phi_3= \doublet{$\begin{scriptsize}$ G^+ $\end{scriptsize}$}{\frac{v+h+iG^0}{\sqrt{2}}}, 
\label{explicit-fields}
\ee
where $\phi_1$ and $\phi_2$ are the two \textit{inert} doublets (odd under the $Z_2$) and $\phi_3$ is the one \textit{active} doublet (even under the $Z_2$) which plays the role of the SM-Higgs doublet, with $h$ being the SM-Higgs boson and $G^\pm,~ G^0$ are the would-be Goldstone bosons.

The Yukawa Lagrangian of the model is identical to the SM Yukawa Lagrangian, with $\phi_3$ playing the role of the 
SM-Higgs doublet:
\bea 
\mathcal{L}_{Yukawa} &=& \Gamma^u_{mn} \bar{q}_{m,L} \tilde{\phi}_3 u_{n,R} + \Gamma^d_{mn} \bar{q}_{m,L} \phi_3 d_{n,R} \nonumber\\
&& +  \Gamma^e_{mn} \bar{l}_{m,L} \phi_3 e_{n,R} + \Gamma^{\nu}_{mn} \bar{l}_{m,L} \tilde{\phi}_3 {\nu}_{n,R} + h.c.  
\eea
where $\Gamma^{u,d,e,\nu }_{mn}$ are the dimensionless Yukawa couplings for the family indices $m,n$
and $u,d,e,\nu$ label the SM fermions in the usual notation.
We assign $Z_2$ charges to each doublet according to the $Z_2$ generator in Eq. (\ref{generator}): odd-$Z_2$ charge to the inert doublets, $\phi_1$ and $\phi_2$, and even-$Z_2$ charge to the active doublet, $\phi_3$. It is clear that the symmetry of the potential is respected by the vacuum alignment $(0,0,\frac{v}{\sqrt{2}})$.

To make sure that the entire Lagrangian and not only the scalar potential is $Z_2$ symmetric, we assign an even $Z_2$ parity to all SM particles, identical to the $Z_2$ parity of the only doublet that couples to them, i.e., the active doublet $\phi_3$. With this parity assignment Flavour Changing Neutral Currents (FCNCs) are avoided as the extra doublets are forbidden to couple to fermions by $Z_2$ conservation.

Note that the scalar $h$ contained in the doublet $\phi_3$ in our model, has exactly the couplings of the SM-Higgs boson. The CP-violation is only introduced in the \textit{inert} sector which is forbidden from mixing with the \textit{active} sector by the $Z_2$ symmetry. Therefore, the amount of CP-violation is not limited by EDMs and SM-Higgs couplings.

The lightest neutral field from the inert doublets which now have a mixed CP-charge, $S_1, S_2, S_3, S_4$, is the DM candidate. To stabilize the DM candidate from decaying into SM particles, we make use of the remnant symmetry of the potential after EWSB \cite{Ivanov:2012hc}. 

Below we study a simplified version of the I(2+1)HDM by imposing the following equalities
\be 
\mu^2_1 =\mu^2_2 , \quad \lambda_3=\lambda_2 , \quad \lambda_{31}=\lambda_{23} ,\quad \lambda'_{31}=\lambda'_{23} 
\ee
which is sometimes referred to as the ``dark democracy'' limit. After imposing this limit, the model is still explicitly CP-violating when $(\lambda_{22}- \lambda_{11} )
\left[\lambda_1 ({\mu^2_{12}}^*)^2-\lambda^*_{1}(\mu^2_{12})^2 \right] \neq 0$ \cite{Haber:2006ue,Haber:2015pua}. Note that in this relation the only parameter that is relevant for our studies is $\mu^2_{12}$ and the rest are ``dark'' parameters which do not play a role in DM or LHC studies.

By imposing the ``dark democracy'' limit, the only two parameters that remain complex are $\mu^2_{12}$ and $\lambda_2$ for which we use the following notation
\bea 
\label{notation}
&&\mu^2_{12} = \Re \mu^2_{12} +i  \Im\mu^2_{12} = |\mu^2_{12}| e^{i \theta_{12}}\\
&&\lambda_2 = \Re \lambda_2 +i  \Im\lambda_2 = |\lambda_2| e^{i \theta_2}. \nonumber
\eea
The angles $\theta_{12}$ and $\theta_2$ are therefore the CP-violating phases of $\mu^2_{12}$ and $\lambda_2$, respectively. 

\subsection{Minimization of the potential}
\label{minimization}

The minimum of the potential sits at the point $(0,0,\frac{v}{\sqrt{2}})$ with
$
v^2= \frac{\mu^2_3}{\lambda_{33}} .
$

\begin{figure}[h!]
\centering
\includegraphics[scale=0.58]{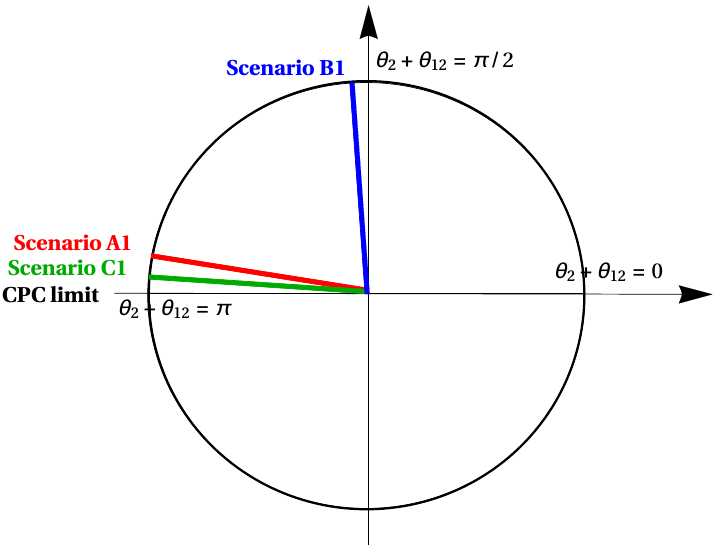}
\caption{The sum of angles $\theta_2+\theta_{12}$ populates the second quadrant. Point $\theta_2+\theta_{12}=\pi$ corresponds to the CP-conserving limit. At the point $\theta_2+\theta_{12}=\pi/2$ the values $\Lambda=\Lambda'$ and mass degeneracies arise where $m^2_{S_1}=m^2_{S_3}$ and $m^2_{S_2}=m^2_{S_4}$. Scenarios A1, B1, C1 chosen for our numerical studies in Section \ref{numerical-analysis} have also been shown here.}
\label{angles}
\end{figure}

\vspace{5mm}
\noindent The mass spectrum of the scalar particles is as follows.
\begin{itemize}
\item \textbf{The fields from the active doublet}\\
The fields from the third doublet, $G^0,G^\pm,h$, which play the role of the SM-Higgs doublet fields have squared masses:
\bea 
&& m^2_{G^0}= m^2_{G^\pm}=0, \nonumber\\
&& m^2_{h}= 2\mu_3^2. 
\eea

\item \textbf{The charged inert fields}\\
The two physical charged states, $S^\pm_1$ and $S^\pm_2$ , from the two inert doublets are the eigenstates of the mass-squared matrix 
\be 
\left(\begin{array}{cc}
-\mu_1^2 +\frac{1}{2}\lambda_{31}v^2 & -\Re\mu_{12}^2+i\Im\mu_{12}^2 \\[2mm]
-\Re\mu_{12}^2-i\Im\mu_{12}^2 & -\mu_2^2 +\frac{1}{2}\lambda_{23}v^2 \\
\end{array}
\right)
\ee
with masses
\be  
m^2_{S^\pm_{1}} = 
(-\mu_2^2 - |\mu_{12}^2|) + \frac{1}{2} \lambda _{23} v^2 , \quad
m^2_{S^\pm_{2}} =
(-\mu_2^2 + |\mu_{12}^2|)  + \frac{1}{2} \lambda _{23} v^2 .
\ee 
The gauge eigenstates can be written in terms of the mass eigenstates:
\be 
H^\pm_{1}= 
\frac{e^{\pm i \theta_{12}/2}}{\sqrt{2}}(S^\pm_{1} - S^\pm_{2}) , 
\qquad 
H^\pm_{2}= \frac{e^{\mp i \theta_{12}/2}}{\sqrt{2}} (S^\pm_{1} + S^\pm_{2}).
\ee

\item \textbf{The CP-mixed neutral inert fields}\\
The four neutral physical states of mixed CP in the basis of $(H^0_1,H^0_2,A^0_1,A^0_2)$ are the eigenstates of the following mass-squared matrix, $\mathcal{M}$:
\begin{footnotesize}
\be 
\label{neutral-mass-squared-matrix}
\mathcal{M}=\left(\begin{array}{cccc}
a
& c
& e 
& -d \\[2mm]
c
& a
& d
& -e \\[2mm]
e
& d
& b   
& c   \\[2mm]
  -d
& -e 
&  c
&  b  \\
\end{array}
\right) 
\ee
\end{footnotesize}
with
\bea 
&& a  =-\frac{\mu_2^2}{2} +(\frac{\lambda_{23}+\lambda'_{23} + 2|\lambda_2|\cos\theta_2}{4})v^2, \quad
b=-\frac{\mu_2^2}{2} +(\frac{\lambda_{23}+\lambda'_{23}-2|\lambda_2|\cos\theta_2}{4})v^2 \nonumber\\
&& c=-\frac{|\mu^2_{12}|\cos\theta_{12}}{2} ,\quad 
d= -\frac{|\mu_{12}^2|\sin\theta_{12}}{2} ,\quad
e= -\frac{v^2|\lambda_{2}|\sin\theta_{12}}{2}. \nonumber
\eea

%
%

The masses of the neutral inerts are
\bea
\label{masses-Ss}
m^2_{S_1} &=&  
\frac{v^2}{2}(\lambda'_{23}+\lambda_{23}) -\Lambda-\mu^2_2,
\\
m^2_{S_2} &=&  
\frac{v^2}{2}(\lambda'_{23}+\lambda_{23}) + \Lambda-\mu^2_2, 
 \nonumber\\
m^2_{S_3} &=& 
\frac{v^2}{2}(\lambda'_{23}+\lambda_{23}) -\Lambda' -\mu^2_2, 
 \nonumber\\
m^2_{S_4} &=&  
 \frac{v^2}{2}(\lambda'_{23}+\lambda_{23}) +\Lambda' -\mu^2_2,
 \nonumber
\eea

where
\bea 
\label{lambdas}
\Lambda &=&\sqrt{v^4|\lambda_2|^2+|\mu^2_{12}|^2-2v^2|\lambda_2| |\mu^2_{12}|\cos(\theta_{12}+\theta_2)}, 
\\[2mm]
\Lambda' &=&\sqrt{v^4|\lambda_2|^2+|\mu^2_{12}|^2+2v^2|\lambda_2| |\mu^2_{12}|\cos(\theta_{12}+\theta_2)}. \nonumber
\eea

We require for $S_1$ to be the DM candidate which for a positive $\Lambda,\Lambda'$ leads to $\Lambda'< \Lambda$ which in turn leads to  $\theta_2+\theta_{12}$ to sit in the second quadrant\footnote{For negative $\Lambda,\Lambda'$, simply the order of the neutral inert particles is changed. The phenomenology of the model is the same by keeping $\theta_2+\theta_{12}$ in the second quadrant and relabeling the particles.} (see Figure \ref{angles}). We also require $\Re\lambda_2<0$ for the model to recover the results in \cite{Keus:2014jha,Keus:2015xya} in the CP-conserving limit. All other parameters are assumed to be positive.

The mass eigenstates can be written in terms of the gauge eigenstates
\bea 
&& 
S_1 =\frac{\alpha H_1^0 + \alpha H_2^0-A_1^0+A_2^0}{\sqrt{2\alpha^2+2}},\quad
S_2 =\frac{-H_1^0-H_2^0 -\alpha A_1^0+ \alpha A_2^0}{\sqrt{2\alpha^2+2}},  \\
&& 
S_3 =\frac{\beta H_1^0 -\beta H_2^0+A_1^0+A_2^0}{\sqrt{2\beta^2+2}},
\quad
S_4 =\frac{- H_1^0+ H_2^0 +\beta A_1^0 +\beta A_2^0}{\sqrt{2\beta^2+2}}, \nonumber
\eea
with
\be
\label{alpha-beta} 
\alpha 
= 
\frac{
-| \mu^2_{12}|  \cos\theta_{12}
+v^2 | \lambda_2|  \cos\theta_2
-\Lambda}{| \mu^2_{12}|  \sin\theta_{12}+v^2 | \lambda_2|  \sin\theta_2}
,\qquad
\beta = \frac{
| \mu^2_{12}|  \cos\theta_{12}
+v^2 | \lambda_2|  \cos\theta_2
-\Lambda'}{| \mu^2_{12}|  \sin\theta_{12}-v^2 | \lambda_2|  \sin\theta_2}.
\ee

\end{itemize}

It is useful to write the parameters of the model in terms of the physical observables:
\bea
\label{parameters}
&& |\mu^2_{12}| = \frac{1}{2}(m^2_{S^\pm_2} -m^2_{S^\pm_1}) ,
\\
&&
\lambda_{23}=\frac{2\mu^2_2}{v^2}+\frac{m^2_{S^\pm_2} +m^2_{S^\pm_1}}{v^2},
\nonumber\\
&&
\lambda'_{23}=\frac{1}{v^2}(m^2_{S_2}+m^2_{S_1}-m^2_{S^\pm_2}-m^2_{S^\pm_1}), \nonumber\\
&&
\mu^2_2= \frac{v^2}{2}g_{S_1S_1h} -\frac{v^2|\lambda_2|}{2(1+\alpha^2)}
\biggl(
4\alpha \sin\theta_2  +2(\alpha^2-1)\cos\theta_2 
\biggr)
-\frac{m^2_{S_2}+
m^2_{S_1}}{2},  \nonumber\\
&&
|\lambda_2|= \frac{1}{v^2}
\left[|\mu^2_{12}|\cos(\theta_2 +\theta_{12})+\sqrt{|\mu^2_{12}|^2 \cos^2(\theta_2 +\theta_{12})+ \left(\frac{m^2_{S_2}-m^2_{S_1}}{2}\right)^2-|\mu^2_{12}|^2} ~\right].
\nonumber
\eea
We take the masses of $S_{1,2},S^\pm_{1,2}$, the two angles $\theta_2$ and $\theta_{12}$ and the Higgs-DM coupling, $g_{S_1S_1h}$ (with the Lagrangian term equal to $\frac{v}{2}g_{S_1 S_1h} h S_1^2$) as the input parameters of the model.

\subsection{Recovering the CPC limit}

In the CP-conserving limit, the purely CP-even particle $H_1$ is assumed to be the DM candidate for which $\lambda_2<0$ \cite{Keus:2014jha,Keus:2015xya}. It can be seen from Eq. (\ref{notation}) that this limit can be recovered by taking $\theta_2=\pi$ and $\theta_{12}=0$.

With $\theta_2+\theta_{12}=\pi$ and $\cos(\theta_2+\theta_{12})=-1$ the values of $\Lambda$ and $\Lambda'$ reduce to
\be 
\Lambda =v^2|\lambda_2|+|\mu^2_{12}|, \qquad 
\Lambda' = v^2|\lambda_2| -|\mu^2_{12}|
\ee 
and the $\alpha$ and $\beta$ parameters tend to infinity resulting in $S_1$ turning into a purely CP-even state with the Higgs-DM coupling
\bea 
g^{CPV}_{hDM} &=& 
\frac{1}{1+\alpha^2}
\biggl[
4\alpha \Im\lambda_2 +2(\alpha^2 -1)\Re\lambda_2 \biggr]+
\lambda_{23} + \lambda'_{23} 
\\[2mm]
&\rightarrow& 2\lambda_2 + \lambda_{23} + \lambda'_{23} = g^{CPC}_{hDM}. 
\eea


%

\newpage
\section{Constraints on parameters}\label{constraints}

\subsection{Theoretical constraints}
In the ``dark democracy" limit, theoretical requirements of boundedness of the potential and positive-definiteness of the Hessian put the following constraints on the potential.
\begin{enumerate}
\item
\textbf{Boundedness of the potential}
\\For the $V_0$ part of the potential to have a stable vacuum (bounded from below) the following conditions are required\footnote{These conditions are resulted from requiring the quartic part of the potential to be positive as the fields $\phi_i \to \infty$. The ``copositivity" method suggested in \cite{Kannike:2012pe} will result in less restrictive constrains.}:
\bea
&& \bullet\quad \lambda_{11}, \lambda_{22}, \lambda_{33} > 0 \\
&& \bullet\quad \lambda_{12} + \lambda'_{12} > -2 \sqrt{\lambda_{11}\lambda_{22}} \nonumber\\
&& \bullet\quad \lambda_{23} + \lambda'_{23} > -2 \sqrt{\lambda_{22}\lambda_{33}} \nonumber
\eea
We also require the parameters of the $V_{Z_2}$ part to be smaller than the parameters of the $V_0$ part:
\be 
\bullet\quad |\lambda_1|, |\lambda_2| < |\lambda_{ii}|, |\lambda_{ij}|, |\lambda'_{ij}| , \quad i\neq j = 1,2,3.
\ee

\item
\textbf{Positive-definiteness of the Hessian}
\\For the point $(0,0,\frac{v}{\sqrt{2}})$ to be a minimum of the potential, the second order derivative matrix must have positive definite determinant. Therefore, the following constraints are required:
\bea 
&& \bullet\qquad   \mu^2_3  > 0   \\
&& \bullet\quad  
\biggl(-\mu _2^2 + (\lambda_{23}+\lambda'_{23})\frac{v^2}{2}\biggr)^2  > | \mu^2_{12} |^2 \nonumber
\eea

\item 
\textbf{Positivity of the mass eigenstates}\\
Further constrains on the parameters of the potential are achieved by requiring the mass eigenstates in each case to be positive:

\bea 
&& \bullet ~~ 
\frac{v^2}{2}(\lambda'_{23}+\lambda_{23}) \pm \Lambda-\mu^2_2 >0
\\
&&  \bullet ~~
\frac{v^2}{2}(\lambda'_{23}+\lambda_{23}) \pm \Lambda' -\mu^2_2 >0 
 \nonumber\\
&&  \bullet ~~ 
(-\mu_2^2 \pm |\mu_{12}^2|) + \frac{1}{2} \lambda _{23} v^2 >0 \nonumber
\eea

\item
\textbf{Meaningful parameters}\\
Extra conditions are required for the expression under the square root in Eqs. (\ref{lambdas} and \ref{parameters}) to be positive

\bea
&&\bullet ~~ 
v^4|\lambda_2|^2+|\mu^2_{12}|^2 \pm 2v^2|\lambda_2| |\mu^2_{12}|\cos(\theta_{12}+\theta_2) > 0 
\\[2mm]
&& \bullet ~~  \left(\frac{m^2_{S_2}-m^2_{S_1}}{2}\right)^2-\left(\frac{m^2_{S^\pm_2} -m^2_{S^\pm_1}}{2}\right)^2 > 0 \nonumber
\eea

As mentioned before, for $S_1$ to be the DM candidate 
\be 
\bullet ~~  \Lambda' < \Lambda ~ \Rightarrow ~ \pi/2 < \theta_2 +\theta_{12} < \pi
\ee
and for $\lambda_2 < 0$ we require
\be 
 \bullet ~~  \pi/2 < \theta_2 < \pi 
\ee

\end{enumerate}

\subsection{Experimental constraints}
\label{experimental}

Properties of all inert scalars, including $S_1$, the DM candidate, are constrained by various experimental results. 

\begin{enumerate}
\item \textbf{Relic density measurements} 

The relic density of $S_1$ is constrained by Planck data \cite{Ade:2015xua}:
\be 
\Omega_{\rm DM} h^2 = 0.1199 \pm 0.0027. \label{PLANCK_lim}
\ee
If $S_1$ constitutes 100\% of DM  in the Universe, then its relic density should lie within the above bound. A DM candidate with $\Omega_{\rm DM} h^2 $ smaller than the observed value is allowed, however, an additional DM candidate is needed to complement the missing relic density. Regions of the parameter space corresponding to values of $\Omega_{\rm DM}h^2$ larger than the Planck upper limit are excluded.

\item  \textbf{Gamma-ray searches} 

Indirect detection experiments measure the product of DM annihilation or decay with respect to the standard astrophysical sources. Especially important here are the measurements of the photon spectra, originating either from the so-called soft channels (quark and boson final states) and hard channels (lepton pairs). The non-detection of a significant excess of photons over the expected astrophysical background places strong constraints on DM mass and its coupling to the visible sector. For the light DM, which is annihilating into $bb$ or $\tau\tau$, the strongest constraints come from the Fermi-LAT satellite, ruling out the canonical cross section 
$\langle \sigma v\rangle \approx 3\times 10^{-26}~{\rm cm}^3/{\rm s}$ for $m_{\rm DM} \lesssim 100 \mbox{ GeV}$ 
\cite{Ackermann:2015zua}.

For the heavier DM candidates the  PAMELA and Fermi-LAT experiments provide similar limits of 
$ \langle \sigma v\rangle \approx 10^{-25}~{\rm cm}^3/{\rm s}$ for  $m_{\rm DM}=200 \mbox{ GeV} $ 
in the $bb,\tau\tau$ or $WW$ channels \cite{Cirelli:2013hv}. HESS measurements of signal 
coming from the Galactic Centre set limits of $ \langle \sigma v\rangle \approx 10^{-25}-10^{-24}~{\rm cm}^3/{\rm s}$ 
for masses up to TeV scale \cite{Abramowski:2011hc}. 

\textbf{Monochromatic gamma lines} 

Further constrains for DM mass and properties could come from the observation of a photon line emission from $\gamma \gamma$, $Z \gamma$ or $h \gamma$ final states. As no standard astrophysical processes are known to produce a monochromatic $\gamma$-line emission, a detection of such a signal would constitute a ``smoking gun'' discovery of DM. It should be remembered,  however, that a neutral DM candidate does not couple directly to photons, therefore a possible annihilation and decay into $\gamma \gamma$ is loop-suppressed. In models such as the I(2+1)HDM the strength of this process can be enhanced by a contribution from another charged particle ($S_{1,2}^\pm$) and will depend on the, otherwise unconstrained and not relevant for relic density calculations, self-coupling parameters $\lambda_{11,12,22},\lambda_1,\lambda'_{12}$.

\item \textbf{DM direct detection} 

The current strongest upper limit on the spin independent (SI) scattering cross section of DM particles on nuclei 
$\sigma_{DM-N}$ is provided by the LUX experiment \cite{Akerib:2015rjg,newlux}.
Future bounds will come from XENON1T, relevant for all regions of DM mass \cite{Aprile:2012zx}.

\item \textbf{Gauge bosons width} 

Bounds coming from limits for the total width of the EW  gauge bosons \cite{Agashe:2014kda} constrain the masses of the inert scalars:
\begin{equation}\label{eq:gwgz}
m_{S_i,S_j}+m_{S_{1,2}^\pm}\,\geq\,m_W,\,m_{S_i}+m_{S_j}\,\geq\,m_Z,\,2\,m_{S_{1,2}^\pm}\,\geq\,m_Z, i,j=1,2,3,4.
\end{equation} 

\item \textbf{Charged scalars} 

A conservative lower limit for the mass of charged scalars \cite{Pierce:2007ut} si taken to be:
 $m_{S_{1,2}^\pm}\,\geq\,70\,\GeV$.

\item \textbf{Collider searches} 

We adopt the limits for the IDM derived from the collider searches for DM, based on the  reinterpretation of LEP 
and LHC run I analyses \cite{Lundstrom:2008ai,Belanger:2015kga}, thereby excluding a region where simultaneously:
\begin{equation}\label{eq:leprec}
m_{S_i}\,\leq\,100\,\GeV,\,m_{S_1}\,\leq\,80\,\GeV,\,\, \Delta m {(S_1,S_i)}\,\geq\,8\,\GeV, i=2,3,4.
\end{equation}

 \item \textbf{Lifetime of charged scalars} 
 
 In order to evade bounds from long-lived charged particle searches, an upper limit for the lifetime of charged scalars is set to be $\tau\,\leq\,10^{-7}$ s, to guarantee their decay within the detector. This translates to an upper bound on the total decay width of the charged scalars $S_{1,2}^\pm$ of $\Gamma_\text{tot}\,\geq\,6.58\,\times\,10^{-18}\,\GeV$. In the studied benchmarks typically the mass of both charged scalars is above 100 GeV and their decay width, driven by $S^{\pm}_i \to S_j W^\pm$, is of the order of $10^{-1}$ GeV, well within the chosen limit.
 
\item \textbf{Invisible Higgs decays}  

The total Higgs decay width in the I(2+1)HDM can be significantly modified with respect to the SM if $h$ can decay invisibly into inert particles. Measurements of invisible Higgs decays limit models in which the Higgs boson can decay into lighter particles which escape detection. Current experimental values provided by the ATLAS and CMS experiments and limits from global fits on the Higgs signal strengths 
on the ensuing Branching Ratio (BR) are \cite{invisible, Belanger:2013xza}:
\be
\textrm{Br}(h \to {\rm inv}) <0.23-0.36, \label{invlimit}
\ee
where $h \to {\rm inv}$ represents the SM-Higgs decay to invisible particles channels.

The partial decay width for the invisible channel $h \to S_1 S_1$ is:
\be
\Gamma(h\to S_1 S_1)=\frac{g_{S_1 S_1h}^{2}v^2}{32\pi m_{h}}\left(1-\frac{4m_{S_1}^{2}}{m_{h}^{2}}\right)^{1/2},\label{invdec} 
 \ee
and
\be
\textrm{Br}(h \to {\rm inv}) = \frac{\Gamma(h\to S_1S_1)}{\Gamma_h^{\rm SM}+\Gamma(h \to S_1S_1)}. \label{invs1}
\ee
 
The bound can be applied in a straightforward way if there is only one particle into which the Higgs boson can decay invisibly. However, for certain cases there can be more unstable particles with $m_i < m_h/2$. They can decay at tree-level in the following way (with the mass order $m_{S_1} < m_{S_3} <m_{S_4} < m_{S_2}$\footnote{For $\Lambda'< \Lambda$ Eq. (\ref{masses-Ss}) leads to this mass ordering.}):
\begin{eqnarray}
S_3 \to Z S_1, \quad S_4 \to Z S_1, \quad S_2 \to Z S_{3,4} \to ZZS_1.
\end{eqnarray}
Notice that, although there are $h S_i^+ S_i^-$ vertices, and both $S^\pm_i$ are unstable with a lifetime of the order of $10^{-20}$ s, this decay will not influence the Higgs invisible decays for studied parameter space as $m_{S_i^\pm} > m_h/2$.  

If the lifetime of $S_{2,3,4}$ is low enough ($\tau < 10^{-7}$ s), neutral particles can decay inside the detector and then the Higgs can decay into:
\begin{eqnarray}
&& h \to S_1 S_1 \;(\textrm{invisible decay}) \label{hdec1}\\
&& h \to S_1 S_2 \to S_1 S_1 Z^*Z^* \;(\textrm{missing energy + decay products of {\it Z}}) \label{hdec2}\\
&& h \to S_3 S_4 \to S_1 S_1 Z^*Z^* \;(\textrm{missing energy + decay products of {\it Z}}) \label{hdec6}\\
&& h \to S_3 S_3 \to S_1 S_1 Z^*Z^* \;(\textrm{missing energy + decay products of {\it Z}})\label{hdec3}\\
&& h \to S_4 S_4 \to S_1 S_1 Z^*Z^* \;(\textrm{missing energy + decay products of {\it Z}}) \label{hdec4}\\
&& h \to S_2 S_2 \to S_1 S_1 Z^*Z^*Z^*Z^* \;(\textrm{missing energy + decay products of {\it Z}}) \label{hdec5}
\end{eqnarray}
Then, only the first channel will constitute an invisible decay of the Higgs particle, while in the remaining channels the signature would be missing energy associated with two dilepton pairs from the decay of an off-shell $Z$: $Z^* \to l^+ l^-$. 

If particles $S_{2,3,4}$ are long-lived enough (i.e., with $\Gamma_\text{tot}(S_i)\,\leq\,6.58\,\times\,10^{-18}\,\GeV \Leftrightarrow \tau\,\geq\,10^{-7}$ s), they will not decay inside the detector, and therefore contribute to the Higgs invisible decays $h \to S_i S_i$. The BR  would then be:
\be
\textrm{BR}(h \to {\rm inv}) = \frac{\sum_{i,j,m_{i,j}<m_h/2} \Gamma(h\to S_iS_j)}{\Gamma_h^{\rm SM}+\sum_i \Gamma(h \to S_iS_j)}, \label{inv_all}
\ee
with
\be
\Gamma(h\to S_i S_i)=\frac{g_{h S_i S_i}^{2}v^2}{32\pi m_{h}}\left(1-\frac{4m_{S_i}^{2}}{m_{h}^{2}}\right)^{1/2}
 \ee
 and
\be
\Gamma(h\to S_i S_j)=\frac{g_{h S_i S_j}^{2}v^2}{32\pi m_{h}^3}\left( (m_h^2-(m_{S_i}+m_{S_j})^2)(m_h^2-(m_{S_i}-m_{S_j})^2)\right)^{1/2}.
 \ee
 
However, for all studied cases, the mass splittings, and therefore the decay widths, of $S_{2,3,4}$ are large enough to ensure a decay inside the detector. 


\item \textbf{Higgs total decay width} 

For $m_{S_i} > m_h/2$ the Higgs total decay width is not changed with respect to the SM by the presence of additional particles (neglecting the change in the partial width $h\to \gamma \gamma$). If $m_{S_i} < m_h/2$ the total decay width is augmented by additional decay channels:
\be
 \mu_{tot} = \frac{\text{BR}(h \to XX) }{\text{BR} (h_{\rm SM} \to XX )} = \frac{\Gamma^{SM}_{tot}(h)}{\Gamma^{SM}_{tot}(h)+\Gamma^{inert}(h)} = 1 - {\rm BR}(h\to \sum_{i,j} S_i S_j).
\ee
Following \cite{Agashe:2014kda} we use $\mu_{tot} = 1.17\pm 0.17$ which leads to the limit of 
\be
{\rm BR}(h\to \sum_{i,j} S_i S_j) < 0.34 \label{inv_tot}
\ee 
at 3$\sigma$ level, which is more restrictive than the direct limit of $\Gamma_h<22$ MeV from \cite{Khachatryan:2014iha}.
%
%

\item \textbf{The $h\to \gamma \gamma$ signal strength} 

The signal strength of Higgs decay into two photons limits the contribution from New Physics (NP) to Higgs observables. The current combined limit from ATLAS and CMS for the Higgs decay into $\gamma \gamma$ via the corresponding signal strength is $\mu_{\gamma\gamma}=1.16^{+0.20}_{-0.18}$ \cite{comb}. It is defined with respect to the SM as:

\begin{equation}
 \mu_{\gamma\gamma} = \frac{ \sigma( gg\to h) }{ \sigma( gg\to h_{\rm SM} ) }  
                \frac{\text{BR}(h \to \gamma\gamma) }{\text{BR} (h_{\rm SM} \to \gamma\gamma) } ,
\end{equation}
assuming: (i) the gluon fusion is the dominant Higgs production channel at the LHC, (ii) the narrow-width approximation, (iii) $\sigma( gg\to h ) = \sigma( gg\to h_{\rm SM} )$ as the Higgs-gg loop is not modified with respect to the SM.  The expression for
$\mu_{\gamma\gamma}$ reduces then to:
\begin{equation}\label{Rxx}
 \mu_{\gamma\gamma} = \frac{\Gamma(h\to \gamma\gamma)^{\textrm{3HDM}} \, \Gamma(h)^{\textrm{SM}}}{\Gamma(h\to \gamma\gamma)^{\textrm{SM}} \, \Gamma(h)^{\textrm{3HDM}}} .
\end{equation}
In the 3HDM $\mu_{\gamma\gamma}$ can be modified both by the presence of light neutral scalars, contributing to $\Gamma(h)^{\textrm{3HDM}}$, and by charged scalars, which change $\Gamma(h\to \gamma\gamma)^{\textrm{3HDM}}$.

\begin{itemize}
\item \textbf{Contribution to $\Gamma(h\to \gamma\gamma)^{\textrm{3HDM}}$:} The one-loop coupling of $h$ to photons receives contributions mainly from $W^\pm$, $t$ and two charged scalars $S_{1,2}^\pm$ from the inert sector, so 
the amplitude can be written as:
\begin{equation}
 A(h\to \gamma \gamma) = A^{SM}_W+ A^{SM}_t + A_{S_1^\pm} + A_{S_2^\pm},
 \end{equation}
 where $A^{SM}_W$ and  $A^{SM}_t$ are the SM contribution from $W^\pm$ and the top quark. Notice that  the ``dark democracy'' limit ensures that there is no $h S_i^+ S_j^-$ vertex and the only relevant loop contributions are due to $h S_i^+ S_i^-$.

The amplitudes are defined as:
\begin{eqnarray}
A_{S_i^{\pm}}=  A_{0}\left(\frac{4m_{S_i^{\pm}}^{2}}{m_{h}^{2}}\right),
A_{t}^{SM}=\frac{4}{3} A_{1/2}\left(\frac{4m^{2}_{t}}{m_{h}^{2}}\right),
A_{W}^{SM}= A_{1}\left(\frac{4m^{2}_{W}}{m_{h}^{2}}\right),
\end{eqnarray}
where
\begin{eqnarray}
A_{1/2}(\tau)&=&2\tau\left[1+(1-\tau)f(\tau)\right],\notag\\
A_{1}(\tau)&=&-\left[2+3\tau+3\tau(2-\tau)f(\tau)\right],\\
A_{0}(\tau)&=&-\tau\left[1-\tau f(\tau)\right]\notag
\end{eqnarray}
and 
\begin{eqnarray}
f(\tau) =
\left\{
	\begin{array}{ll}
\arcsin^{2}(1/\sqrt{\tau})  & \mbox{for } \tau\geq1 \\
-\frac{1}{4}\left(\log\frac{1+\sqrt{1-\tau}}{1-\sqrt{1-\tau}}-i\pi\right)^2  & \mbox{for } \tau<1.
	\end{array}
\right.
\end{eqnarray}\\

The partial $h\to \gamma\gamma$ width then reads:
\begin{eqnarray}
 \Gamma(h \rightarrow \gamma\gamma)^{\mathrm{3HDM}}&=&\frac{G_F\alpha^2M_h^3}{128\sqrt{2}\pi^3}\bigg|\frac{4}{3}A_{1/2}\left(\frac{4m_t^2}{m_h^2} \right)+A_1\left(\frac{4m_W^2}{m_h^2} \right)\nonumber\\&& +\sum\frac{g_{hS_i^+S_i^-} v^2}{2m_{S_i^\pm}^2}A_0 \left(\frac{4m_{S_i^\pm}^2}{m_h^2} \right)\bigg|^2\,, \label{Hloop}
\end{eqnarray}
where the first line shows the SM contribution while the second shows the 3HDM contribution from two charged scalars. Notice, that although $g_{hS_1^+S_1^-} = g_{hS_2^+S_2^-}$, the relative contribution from the heavier $S^\pm_2$ is smaller than the one coming from $S_1^\pm$. The maximum contribution from both scalars will arise for cases where $m_{S_{1}^\pm} \approx m_{S_{2}^\pm}$ and when both $S_i^\pm$ are relatively light.

\item \textbf{Contribution to $\Gamma(h)^{\textrm{3HDM}}$:} as discussed in point 9, the Higgs total decay width will be changed by decays into light inert particles if their masses are smaller than $m_h/2$. For $m_{S_i} > m_h/2$, we get $\Gamma^{\rm SM}(h)\approx \Gamma^{\rm 3HDM}(h)$ as we neglect the change in $\Gamma(h\to \gamma \gamma)$.

\end{itemize}

\item \textbf{$S,T,U$ parameters} 

EW precision measurements can provide strong constraints on NP.   In particular, additional particles may introduce important radiative corrections to gauge boson propagators, parametrized by the oblique parameters $S$, $T$ and $U$. These parameters will be influenced by inert particles $S_i^\pm$, $S_i$, which are contributing to the neutral and charged current processes at low energies ($T$), or to neutral current processes at different energy scales ($S$). $U$ is generally small in NP models. 
The latest values of the oblique parameters, determined from a fit with reference mass values of top and Higgs boson $m_{t}=173~\GeV$ and $M_{h}=125~\GeV$ are \cite{Baak:2014ora}:
\begin{equation}
S = 0.05\pm0.11, \quad T = 0.09\pm0.13, \quad U = 0.01\pm0.11.
\end{equation}
Our parameter choices  are compliant with these limits \cite{Dolle:2009fn}.
\end{enumerate}

\section{Relevant DM (co)annihilation scenarios}\label{coannihilation}
The relic density of the scalar DM candidate, $S_1$, after freeze-out is given by the solution of the Boltzmann equation:
\begin{equation}
\frac{d n_{S_1}}{dt} = - 3 H n_{S_1} - \langle \sigma_{eff} v \rangle (n_{S_1}^2 - n^{eq \; 2}_{{S_1}}),
\end{equation}
where the thermally averaged effective (co)annihilation cross section contains all relevant annihilation processes of any $S_i S_j$ pair into SM particles:
\begin{equation}
\langle \sigma_{eff} v \rangle = \sum_{ij} \langle \sigma_{ij} v_{ij} \rangle \frac{n^{eq}_i}{n^{eq}_{S_1}} \frac{n^{eq}_j}{n^{eq}_{S_1}},
\end{equation}
where
\begin{equation}
\frac{n^{eq}_i}{n^{eq}_{S_1}} \sim \exp({-\frac{m_i - m_{S_1}}{T}}).
\end{equation}
Therefore, only processes for which the mass splitting between a state $S_i$ and the lightest $Z_2$-odd particle $S_1$ are comparable to the thermal bath temperature $T$ provide a sizeable contribution to this sum.

The CP-violating I(2+1)HDM studied here shares many features of a Higgs-portal DM model. In a large region of parameter space the most important channel for the DM annihilation is 
\be 
S_1 S_1 \to h_{\rm SM} \to f \bar f
\label{annihilation-1}
\ee
The efficiency of this annihilation channel depends on both the mass of DM and the Higgs-DM coupling. In general, if $m_{\rm DM} < m_h/2$, then one needs a coupling that is relatively large to produce relic density in agreement with Eq.~(\ref{relic}). In this case a small DM-Higgs coupling leads to too large a relic density and results in the overclosure of the Universe. 

Processes with gauge boson products, such as
\be 
S_1 S_1 \to h_{\rm SM} \to V V, \qquad S_1 S_1 \to V V,
\label{annihilation-2}
\ee
 also contribute to the total annihilation cross section, where $V$ is any of the SM gauge bosons.
Contribution from these processes is suppressed when the DM mass is smaller than $m_W$, however, as studies have shown, diagrams with off-shell gauge bosons may be very important for $m_{\rm DM} < m_W$ in models such as the CP-violating I(2+1)HDM. In our analysis such processes, 
\be 
S_1 S_1 \to V V^* \to V f \bar f, \qquad S_1 S_1 \to V^* V^* \to f\bar f f \bar f,
\label{annihilation-3}
\ee
are also included.

Coannihilation effects play an important role in scenarios with multiple particles that are close in mass. Particles up to 20\% heavier than the DM candidate may influence the DM relic density. Therefore, the coannihilation processes, such as 
\be 
S_1 S_i \to h_{\rm SM} \to f\bar f, \qquad S_1 S_i \to Z^* \to f \bar f, \qquad S_1 S^\pm_j \to W^{\pm *} \to f f'
\label{annihilation-4}
\ee
with $i=2,3,4$ and $j=1,2$ which appear in our analysis
are included in calculating the effective annihilation cross section.

If all inert particles are very close in mass then all following channels 
\be 
S_i S_j \to h_{\rm SM} \to f\bar f, \qquad S_i S_j \to V V
\label{annihilation-5}
\ee
contribute to the final DM relic density.

Taking all such processes into account, relevant DM (co)annihilation cases in the CP-violating I(2+1)HDM are presented in the following benchmark scenarios, in the low and medium mass regions ($m_{S_1} < m_Z$).

\begin{itemize}
\item \textbf{Scenario A }\\
with large mass splittings between the DM candidate and all other inert particles:
\be 
m_{S_1} \ll m_{S_2}, m_{S_3}, m_{S_4}, m_{S^\pm_1}, m_{S^\pm_2}.
\ee
In this scenario no co-annihilation channels are present.

\item \textbf{Scenario B}\\
with a small mass splitting between the DM and only one inert neutral particle,
\be 
m_{S_1} \sim m_{S_3} \ll m_{S_2},  m_{S_4}, m_{S^\pm_1}, m_{S^\pm_2}.
\ee
In this scenario the DM can coannihilate with its only particle close in mass, $S_3$.

\item \textbf{Scenario C}\\
with all neutral particles close in mass:
\be 
m_{S_1} \sim m_{S_3} \sim m_{S_2} \sim  m_{S_4} \ll m_{S^\pm_1}, m_{S^\pm_2}.
\ee
In this scenario the DM can coannihilate with all other neutral inert particles.
\end{itemize}

In the heavy mass region ($m_{S_1}> 400$ GeV), neutral and charged inert particles could be close in mass (see point 5 in Section \ref{experimental}).

\begin{itemize}
\item \textbf{Scenario G}\\
with two separate ``families'' of inert particles, each consisting of one charged scalar and two neutral particles 
where ``one family" of inert particles are close in mass and decoupled from the ``second family" of inert particles
\be 
m_{S_1} \sim m_{S_3} \sim  m_{S^\pm_1} \ll m_{S_2}\sim  m_{S_4}\sim m_{S^\pm_2}.
\ee

\item \textbf{Scenario H}\\
where all inert particles are close in mass
\be 
m_{S_1} \sim m_{S_3} \sim m_{S_2} \sim  m_{S_4} \sim m_{S^\pm_1} \sim m_{S^\pm_2}.
\ee

\end{itemize}

\section{Numerical analysis for chosen benchmarks}
\label{numerical-analysis}
In this Section we present the numerical study of the chosen benchmark scenarios. We focus on three regions of DM mass: the low DM mass region with $m_{S_1} < m_h/2$, the medium DM mass region with $m_h/2 < m_{S_1} < m_Z$ and the heavy DM mass region with $m_{S_1} > 400 \GeV$. Following the discussion in Section \ref{minimization} we have chosen as input parameters four masses, $m_{S_{1,2}}, m_{S^\pm_{1,2}}$, of inert particles and two phases, $\theta_2$ and $\theta_{12}$. It is convenient to introduce the mass splittings between the DM candidate and other inert scalars as:
\be
\delta_{12} = m_{S_2} - m_{S_1}, \; \delta_{1c} = m_{S_1^\pm} - m_{S_1}, \; \delta_{c} = m_{S_1^\pm} - m_{S_2^\pm}.
\ee
We then define three base benchmarks in low and medium mass region as
\begin{eqnarray}
A1: \; \delta_{12} =125 \GeV, \delta_{1c} = 50\GeV, \delta_{c} = 50\GeV, \theta_2 = \theta_{12} =1.5 \\
B1: \; \delta_{12} =125 \GeV, \delta_{1c} = 50\GeV, \delta_{c} = 50\GeV, \theta_2 = \theta_{12} =0.82 \\
C1: \; \delta_{12} =12 \GeV, \delta_{1c} = 100\GeV, \delta_{c} = 1\GeV, \theta_2 = \theta_{12} =1.57
\end{eqnarray}
and two in the heavy DM mass region 
\begin{eqnarray}
G1: \; \delta_{12} = 2\GeV, \delta_{1c} =1 \GeV, \delta_{c} = 1\GeV, \theta_2 = \theta_{12} =0.82 \\
H1: \; \delta_{12} = 50\GeV, \delta_{1c} =1 \GeV, \delta_{c} = 50\GeV, \theta_2 = \theta_{12} = 0.82
\end{eqnarray}

Note that the values of the angles $\theta_2$ and $\theta_{12}$ are chosen to be equal since its only the sum of the angles that plays a role in the DM and LHC phenomenology of the model and not the values of the angles individually.

\subsection{Relation between couplings and DM relic density}

In the CP-conserving version of the I(2+1)HDM (within the ``dark democracy'' limit), couplings between inert scalars and gauge bosons are fixed, and given by the rotation angles $\theta_a=\theta_h=\pi/4$. They do not depend on the mass splittings or the value of $m_{S_1}$. In the CP-violating case the situation is different, as the couplings (normalized to $\frac{i e}{2c_w s_w}$) are given by:
\begin{eqnarray}
&&\chi_{Z S_1 S_3} = \chi_{Z S_2 S_4} = \frac{\alpha+\beta}{\sqrt{\alpha^2+1}\sqrt{\beta^2+1}}, \\
&&\chi_{Z S_1 S_4} = \chi_{Z S_2 S_3} = \frac{\alpha\beta-1}{\sqrt{\alpha^2+1}\sqrt{\beta^2+1}},\\
&&\chi^2_{Z S_{1} S_3} + \chi_{Z S_{1} S_4}^2 = 1
, \quad \chi^2_{Z S_{2} S_3} + \chi_{Z S_{2} S_4}^2 = 1.
\end{eqnarray} 

\begin{figure}[h]
\centering
\includegraphics[scale=1]{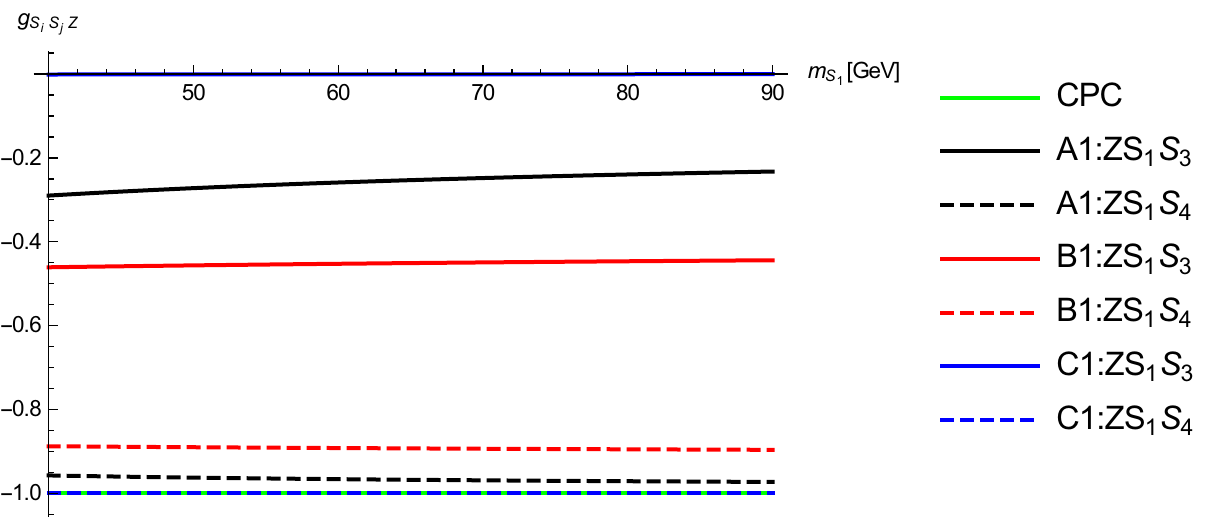}
\caption{Values of $\chi_{Z S_1 S_3} = \chi_{Z S_2 S_4}$ and $\chi_{Z S_1 S_4} = \chi_{Z S_2 S_3}$ couplings for chosen benchmarks. \label{couplingsZ}}
\end{figure}

\begin{figure}[htb]
\vspace{-10pt}
  \centering
  \subfloat[$g_{S_1S_1h} = -0.1$]{\label{couph1}\includegraphics[scale=1]{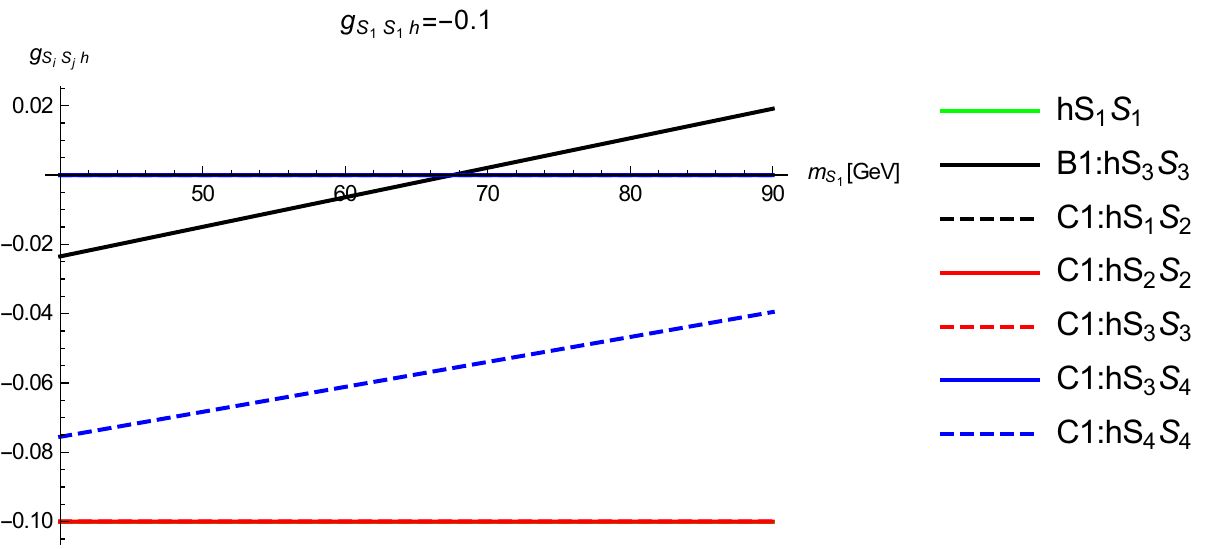}} \\
  \subfloat[$g_{S_1S_1h} = 0.1$]{\label{couph2}\includegraphics[scale=1]{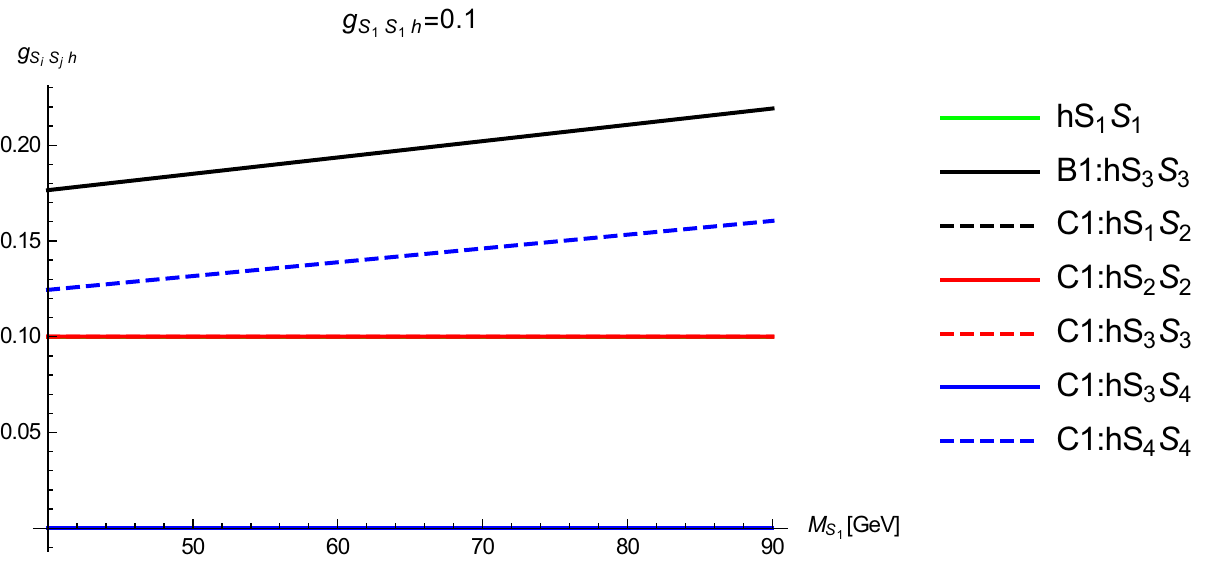}} \\
\vspace{-5pt}
  \caption{Values of the Higgs-inert scalar couplings for chosen benchmarks. \label{couplingsh1}}
\end{figure}

\begin{figure}[htb]
\vspace{-10pt}
  \centering
  \subfloat[$g_{S_1S_1h} = -0.001$]{\label{couph3}\includegraphics[scale=1]{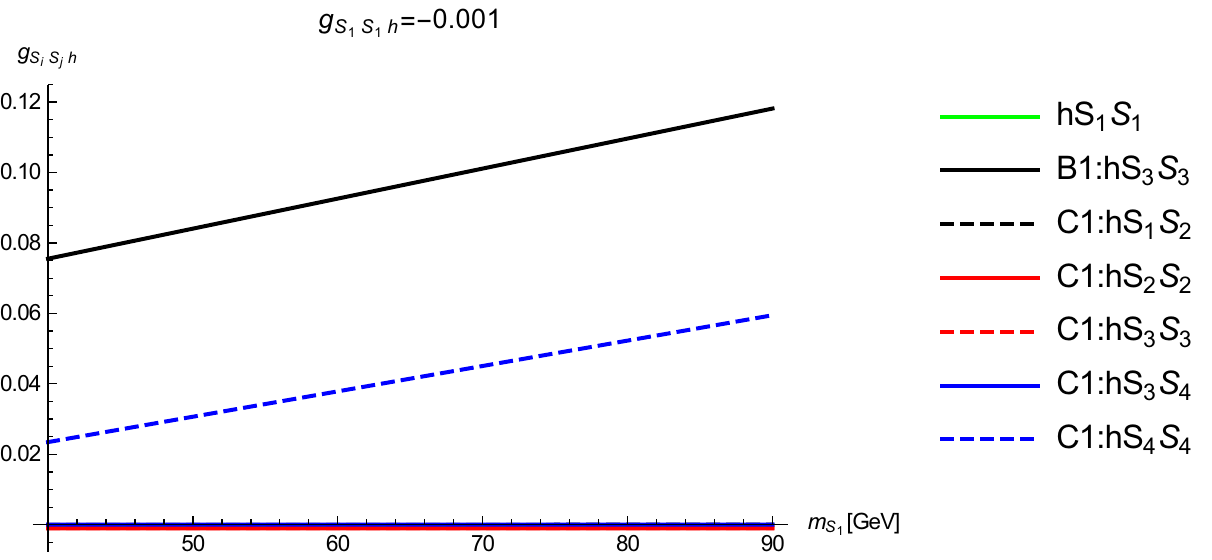}} \\
  \subfloat[$g_{S_1S_1h} = 0.001$]{\label{couph4}\includegraphics[scale=1]{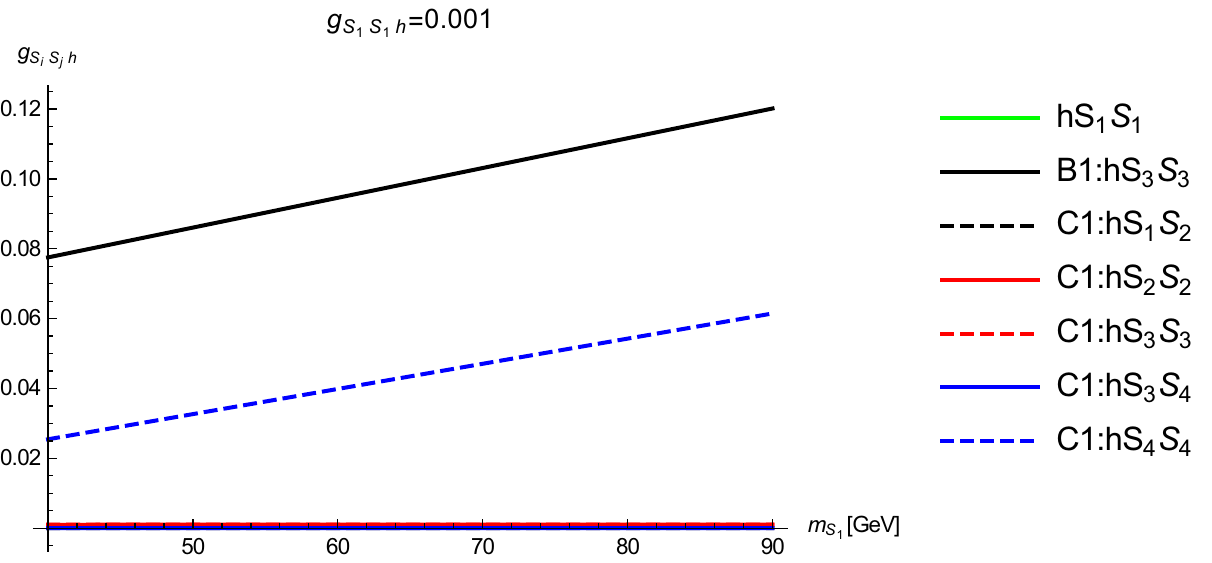}} \\
\vspace{-5pt}
  \caption{Values Higgs-inert scalar couplings for chosen benchmarks. \label{couplingsh2}}
\end{figure}

 The strength of gauge-inert interaction depend on parameters $\alpha$ and $\beta$ in Eq.~(\ref{alpha-beta}), which in turn depend on $m_{S_i}$. Higgs-inert scalar couplings are also modified with respect to the CP-conserving case. This leads to important differences in the DM phenomenology, especially in the region where coannihilation channels are important. Figure (\ref{couplingsZ}) shows the change in values of $Z$-inert couplings for benchmarks A1, B1 and C1, while Figs. \ref{couplingsh1} and \ref{couplingsh2}  present relevant Higgs-inert couplings. The introduction of varying values of $\alpha$ and $\beta$ leads to the following modifications with respect to the (co)annihilation scenarios in the CP-conserving  I(2+1)HDM.

\subsubsection{Low DM mass region}
\begin{enumerate}
\item For benchmark A1, couplings with the $Z$ are modified with respect to the CP-conserving case (Fig. \ref{couplingsZ}), however, as DM does not coannihilate, this change does not modify the annihilation scenario of $S_1$. For low DM mass $S_1$ annihilates mostly through $S_1 S_1 \to h \to b \bar{b}$, entering the resonance region with small Higgs-DM coupling for masses close to $m_h/2$. This benchmark resembles both the CP-conserving I(2+1)HDM as well as the IDM. 

\item For benchmark B1, $S_1$ is close in mass with $S_3$, opening the coannihilation channel $S_1 S_3 \to Z \to f\bar{f}$ (dominant channels with light quarks). Such a scenario in the CP-conserving limit results in too low a relic density for any value of the Higgs-DM coupling due to strong coannihilation between the DM and the next-to-lightest inert particle. In the CP-violating case, however, the strength of the coannihilation channel is reduced. We can therefore change the contribution of this diagram to the relic density calculations not only by introducing the change for the mass splitting, but also by modifying the value of the coupling itself. Diagram $S_1 S_4 \to Z$ is stronger, but because of mass difference this process is not contributing to the relic density calculations. 

We should note that the Higgs-inert couplings change significantly between benchmarks and that they also depend on the value of $m_{S_1}$. In case B especially important is $g_{S_3 S_3 h}$, the coupling of the next-to-lightest inert particle to $h$. Particularly for small values of $g_{S_1 S_1 h}$ it can reach large values and will significantly change the Higgs phenomenology.

\item For benchmark C1 all particles are close in mass and in principle all coannihilation diagrams $S_i S_j \to \textrm{SM} ~\textrm{SM}$ could be important. As the  couplings $g_{S_1 S_2h}, g_{S_3 S_4h}$ and $g_{Z S_1 S_3}$ are suppressed, the crucial contribution comes from $S_1 S_4 \to Z \to q \bar{q}$. In the CP-conserving case, this scenario is only viable in the resonance region. In the CP-violating case, however, the strength of the coannihilation channels depends on the input parameters and can therefore be varied. 

\end{enumerate}

\begin{figure}[h!]
\centering
\includegraphics[scale=1]{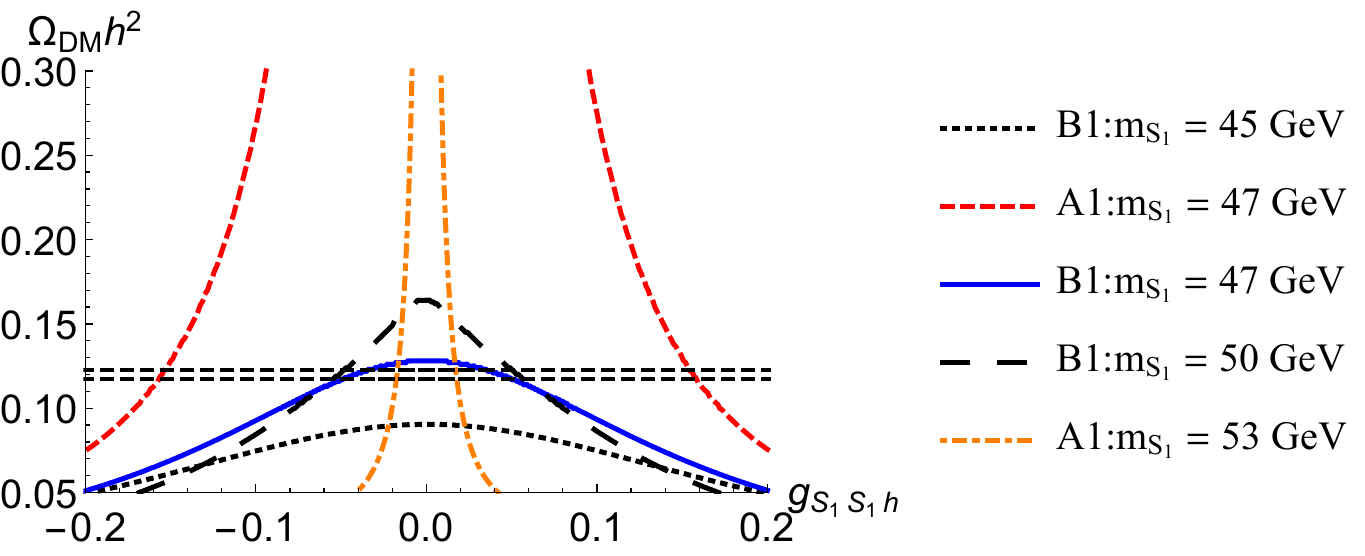}
\caption{Relic density for low DM mass region. The horizonal dashed lines show the Planck limit. \label{omegalow}}
\end{figure} 

To illustrate the varying annihilation scenarios for different parameter choices we have chosen a few points presented in Fig. \ref{omegalow}. 
Scenario A1 with $m_{S_1} = 47 \GeV$ corresponds to the Higgs-portal annihilation into pair $b\bar{b}$, and large coupling is needed to ensure a large enough cross section. As the mass grows, as illustrated by A1 with $m_{S_1} = 53 \GeV$, we are entering the resonance annihilation with suppressed couplings. For case B1, one can see the contribution from coannihiliation channels, that enchance the cross section even for smaller values of coupling. For $m_{S_1} = 45 \GeV$ relic density is too small, however for B1 with $m_{S_1} = 47 \GeV$  it is large enough to fulfil Planck limits. For larger masses, B1 with $m_{S_1} = 50 \GeV$, Higgs-mediated annihilation starts to play a more important role.

\begin{figure}[h]
\centering
\includegraphics[scale=1.1]{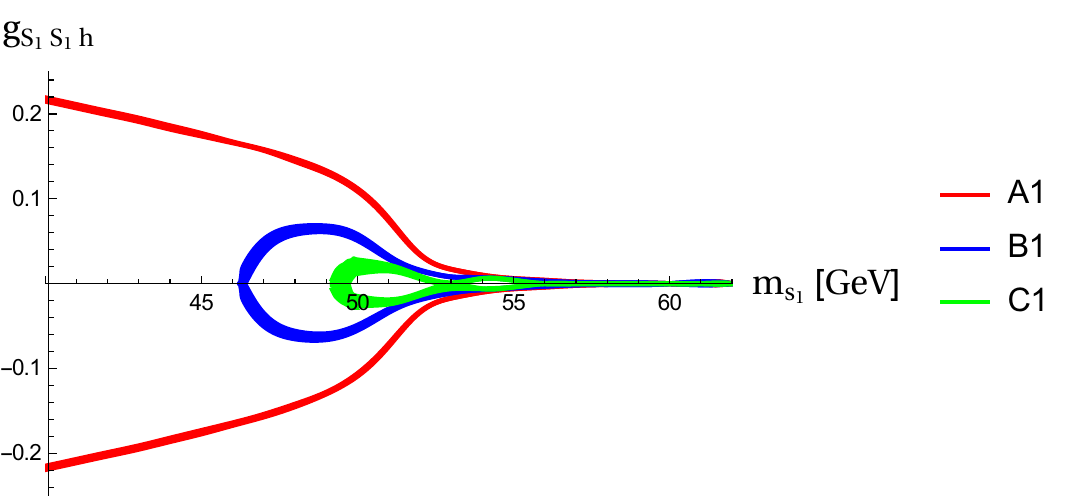}
\caption{Relic density for low DM mass region in Scenarios A1, B1 and C1. \label{relic1}}
\end{figure}

Figure \ref{relic1} shows values of mass and Higgs-DM coupling that produce the correct DM relic density for benchmarks A1, B1 and C1. Benchmark A1 shows the standard behaviour of an $SU(2)$ DM candidate. Benchmark B1, with coannihilation channels, differs from A1. For large values of $g_{S_1S_1h}$ the dominant channel is $S_1S_1 \to \bar{b}b$ and, as there are also coannihilation channels, the relic density is usually too small. For smaller couplings the dominant channel is $S_1 S_3 \to Z \to q \bar{q}$. If the DM mass is small, the relevant cross section is too big. As the mass grows, the coannihilation channel gets weaker, allowing us to obtain the proper relic density. For masses closer to $m_h/2$ the resonance annihilation dominates, following the pattern of benchmark A1. In case of benchmark C1 for small values of $g_{S_1S_1h}$ the dominant channel is $S_1 S_4 \to Z \to f \bar{f}$ (light quarks), with a small contribution from $S_2 S_3 \to Z \to f \bar{f}$. For larger couplings the process  $S_1 S_1 \to h \to b \bar{b}$ strongly increases the annihilation cross section. That, combined with the fact that coannihilation channels are generally strong, leaves the region $m_{S_1} > 49$ GeV.

\subsubsection{Medium DM mass region}

\begin{figure}[h!]
\centering
\includegraphics[scale=1]{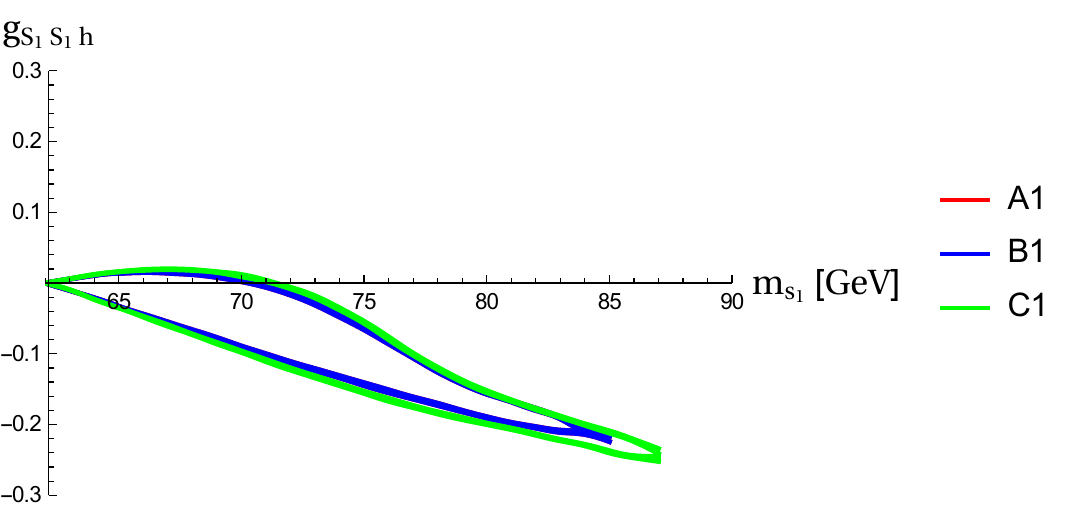}
\caption{Relic density for medium DM mass region in Scenarios A1, B1 and C1. Note that the medium mass region behaviour of the three scenarios is very similar to each other. \label{relic2}}
\end{figure}

\begin{figure}[h!]
\centering
\includegraphics[scale=1]{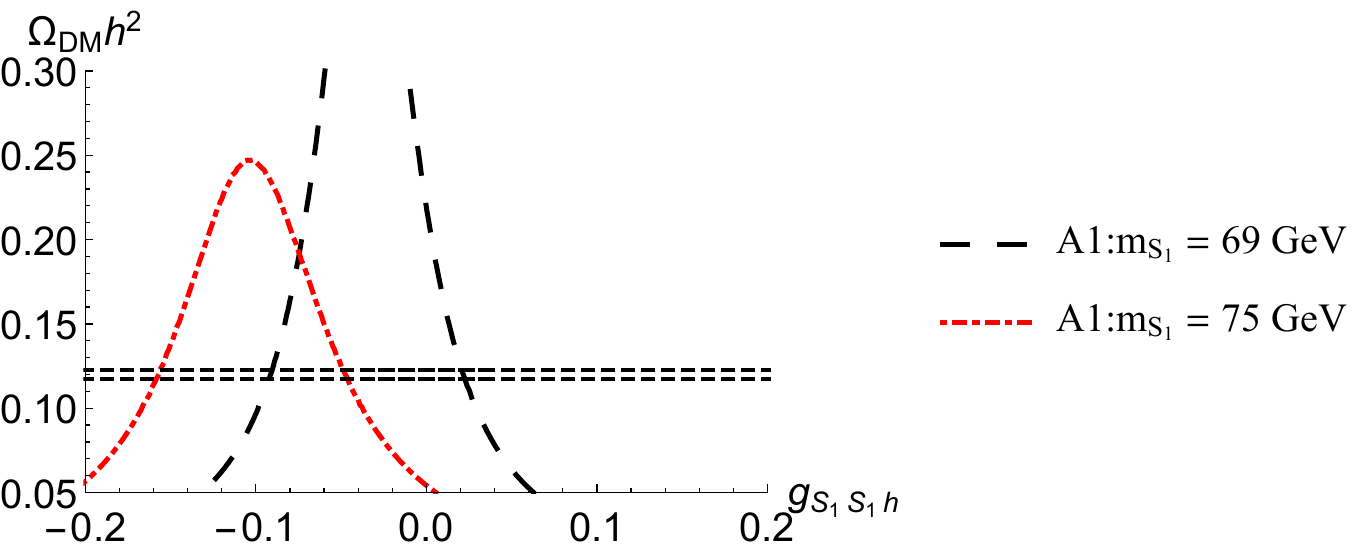}
\caption{Relic density for low DM mass region. The horizonal dashed lines show the Planck limit. \label{omegamid}}
\end{figure}

In the medium DM mass, for $m_h/2 < m_{S_1} < m_{W^\pm,Z}$ the crucial channel for all benchmarks (apart from masses close to $m_h/2$ which are still available following the Higgs-resonance annihilation) is the point annihilation of $S_1 S_1 \to W^+ W^-$ and this vertex does not depend on parameters $\alpha$ and $\beta$. This is the reason, why all studied benchmarks as well as the CP-conserving scenarios follow the similar behaviour, presented in Fig. \ref{relic2}. For larger values of DM mass this annihilation is stronger, and cancellation with $S_1 S_1 \to h \to W^+ W^-$ is needed to ensure the proper value of relic density. This mechanism is responsible for moving towards the negative values of Higgs-DM coupling. Fig. \ref{omegamid} presents two chosen points for benchmark A1, with $m_{S_1} = 69 \GeV$ and $m_{S_1} = 75 \GeV$. In the first case, contribution from $S_1 S_1 \to h \to b \bar{b}$ is still important, while in the second there are mainly gauge boson final states.

In benchmarks B1 and C1 other channels, like $S_1 S_4 \to q \bar{q}$ or $S_3 S_3 \to W^+ W^-$ give small contributions, leading to small deviations from the behaviour of benchmark A1.

\subsubsection{Filling the plot in low and medium mass region}

In the discussion above we have presented results for three sets of parameters in scenarios A1, B1, and C1. It is clear that by changing the input set we can reach different regions of parameters space. Compare, for example, scenarios A1 and B1, which differ only by the chosen values of the sum of $\theta_2$ and $\theta_{12}$. The performed scan shows that by varying the mass splitting and phases $\theta_2$ and $\theta_{12}$ we can actually fill the empty regions in plots \ref{relic1} and \ref{relic2} within the range given by the CP-conserving scenario with large mass splittings (no coannihilation channels). We have more freedom in the low mass region - this is because in the standard CP-conserving case the main annihilation channel is the Higgs-mediated annihilation into $\bar{b}b$. It is easy to obtain strong gauge coannihilation channels. In the medium mass region there is already a strong base annihilation of $S_1$ into $WW$ pair (both direct and Higgs-mediated) and therefore the coannihilation processes have smaller impact.

\begin{figure}[htb]
\centering
\includegraphics[scale=1.2]{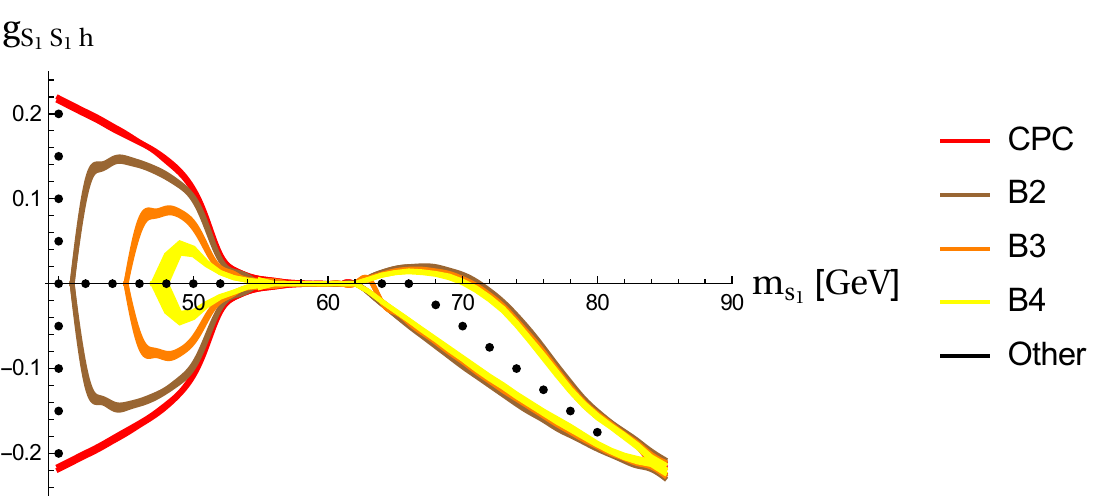}
\caption{The relic density plots for different B and C scenarios where by changing the angles $\theta_2$ and $\theta_{12}$ the whole region not accesible by the CP-conserving limit could be realised in the CP-violating case.\label{filling}}
\end{figure}

In Fig. \ref{filling} results obtained for various additional sets of parameters are presented. We can fill the plot by different B scenarios, where the coannihilation channel $S_1 S_3 \to Z \to q \bar{q}$ (with varying $\chi_{Z S_1 S_3}$ is crucial). It is also possible to find solutions of type C, where all neutral particles have a relatively low mass.

\subsubsection{Heavy DM mass region}

In the heavy DM mass regime necessary ingredients for obtaining a correct value of DM relic density are cancellations between pure gauge and Higgs-mediated annihilation of DM particle, combined with coannihilation channels of, at least, two other scalar particles. Following the analysis for the CP-conserving version of I(2+1)HDM we study two separate scenarios, G1 and H1. 

\begin{figure}[htb]
\centering
\includegraphics[scale=1]{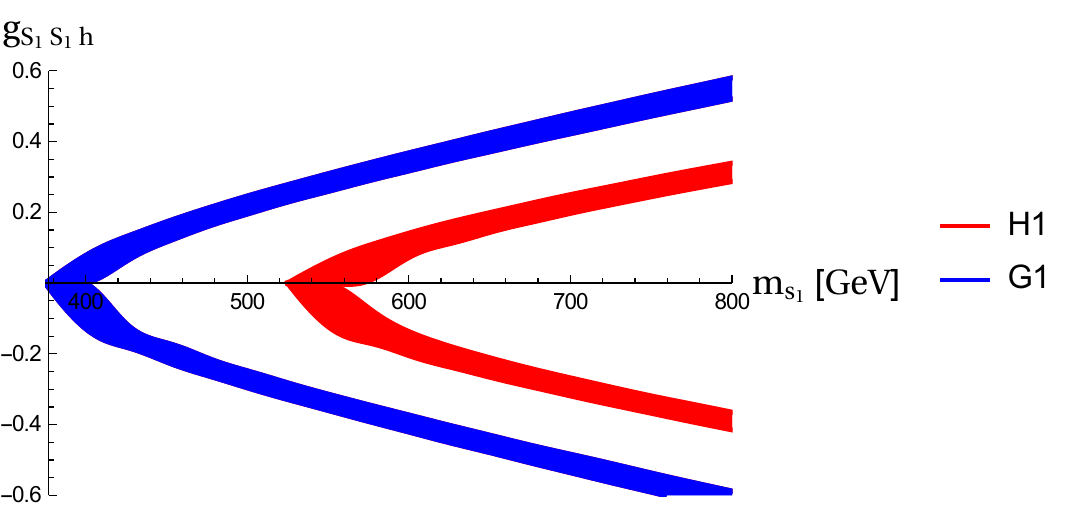}
\caption{Relic density for heavy mass region. \label{relic3}}
\end{figure}

The main (co)annihilation channels are 
\begin{eqnarray}
S_i S_i \to W^+ W^-, ZZ, \quad S_i S_i \to h \to W^+ W^-, ZZ, \\
S_j^\pm S_j^\pm \to W^\pm W^\pm, ZZ, \quad S_j^\pm S_j^\pm \to h \to W^\pm W^\pm,
\end{eqnarray}
where $i=1,3,j=1$ for case H1 and $i=1,2,3,4, j=1,2$ for case G1. We remind the reader that these channels do not depend on $\alpha$ and $\beta$. Dependence on parameters $\alpha$ and $\beta$ appears in the mixed channels, e.g. $S_1 S^\pm_1 \to \gamma W^\pm$. However, these are generally weaker and their influence on the heavy DM relic density studies is minimal. This leads to the known behaviour (as in the CP-conserving case) of the heavy DM candidate, presented in Fig. \ref{relic3}.

\subsection{DM detection experiments}
\subsubsection{DM direct detection}

\begin{figure}[h!]
\centering
\includegraphics[scale=1]{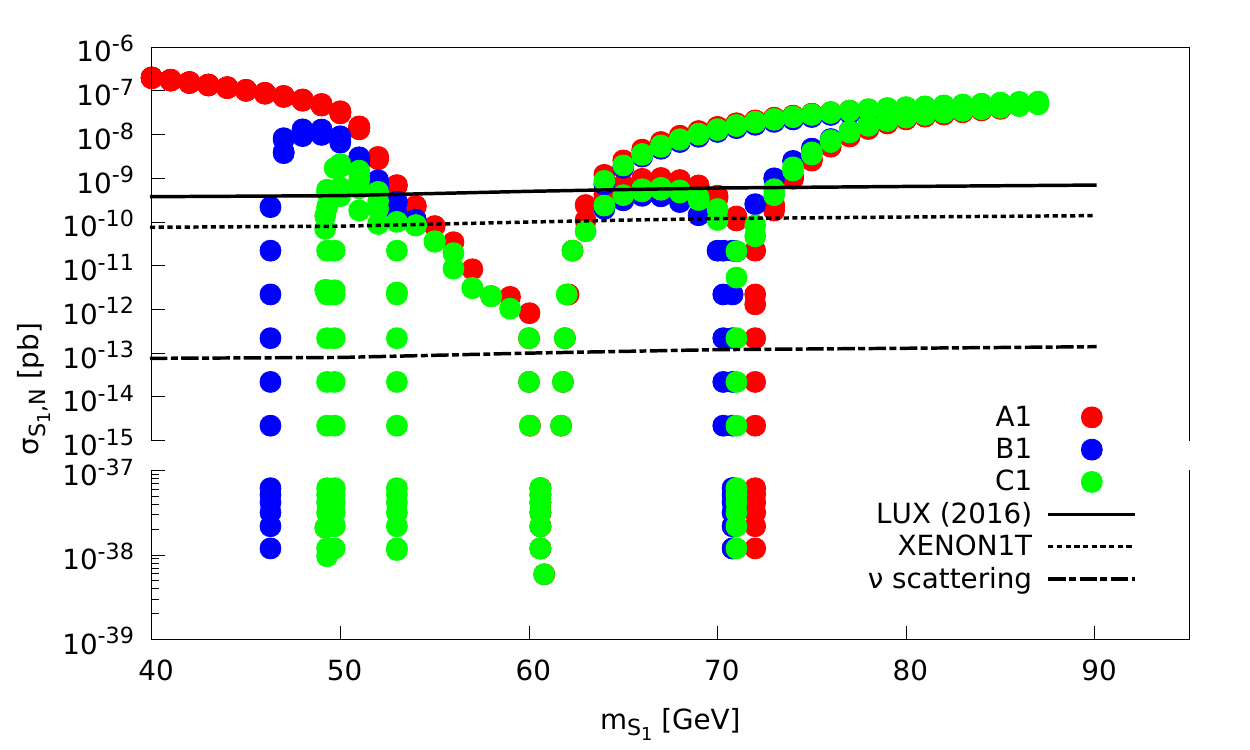}
\caption{Direct detection limits for low and medium mass regions. \label{direct1}}
\end{figure}

DM detection experiments aim to measure the scattering of DM particle off nuclei. This interaction is mediated by the Higgs particle, and therefore results of these experiments constrain the DM mass, as well as its coupling to $h$, following:
\be
\sigma_{DM,N} \propto \frac{g_{S_1 S_1 h}^2 \mu^2 m_N^2}{m_h^4 m_{S_1}^2}, \label{sigmadet}
\ee
where $m_N$ is the nucleon mass and $\mu = m_N m_{S_1}/(m_N + m_{S_1})$ is the reduced nucleon mass. The proportionality constant is given by the square of a matrix element $f_N = 0.30\pm0.03$. In the low and medium mass region the strongest constraints come from the LUX experiment, and they set strong limits on the parameter space of the 3HDM. Results are presented in Fig. \ref{direct1}, where the solid line corresponds to the current LUX limit, while the dashed line shows the projected sensitivity of XENON1T. 

From the plot we can see that for chosen benchmark points A1, B1 and C1 the only surviving region of this part of parameter space is $50 \GeV \lesssim m_{S_1} \lesssim 76 \GeV$. For smaller masses the Higgs-DM coupling needed to obtain good relic density by enhancing the $S_1 S_1 \to h \to b \bar{b}$ channel is too big. For larger masses the coupling needed to cancel the strong annihilation into gauge bosons is generally too big. Two branches in Fig. \ref{direct1} in the medium mass region correspond to two asymmetrical regions from Fig. \ref{relic2}. They do overlap in the low mass region, where good relic density regions from Fig. \ref{relic1} are symmetrical, following relation \ref{sigmadet}. 

\begin{figure}[h!]
\centering
\includegraphics[scale=1]{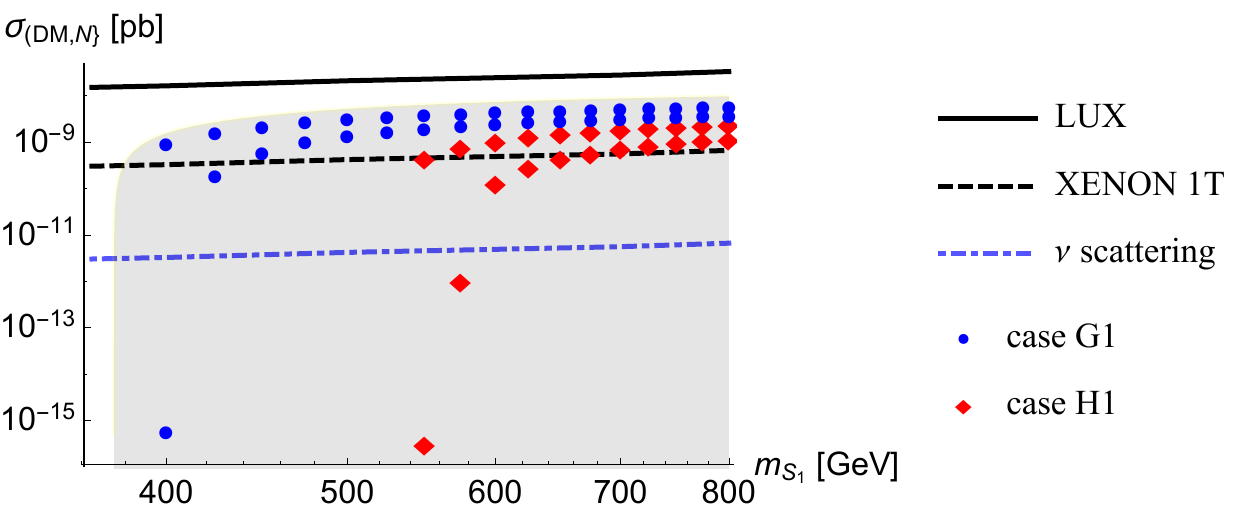}
\caption{Direct detection limits for heavy mass region. \label{direct2}}
\end{figure}

Sensitivity of direct detection experiments drops significantly when applied to heavier DM candidates. Results of the scan for our benchmarks G1 and H1 are presented in fig. \ref{direct2}, where the shaded region corresponds to the probed phase space of the I(2+1)HDM for various choices of mass splittings.

Figs. \ref{direct1}  and \ref{direct2} also shows a  limit from the future XENON1T experiment \cite{Aprile:2012zx}. We expect the next generation of DM detectors, such as XENON1T, to be able to test a large portion of the parameter space of the I(2+1)HDM for $m_{S_1} \lesssim 1$ TeV.

In all regions of DM mass there are points in the parameter space where the Higgs-DM coupling is tending towards zero. It happens in the resonance region for the light DM particle, as well in the heavy mass region for various values of masses, which is related to the cancellation between diagrams. In the heavy mass region  with varying mass splittings it is possible to obtain solutions that require $g_{S_1 S_1 h} \approx 0$. These points will not be tested by the direct detection experiments, as the scattering cross section lies within the coherent neutrino-nucleus scattering regime \cite{Anderson:2011bi} .

\subsubsection{DM indirect detection}

\begin{figure}[h!]
\centering
\includegraphics[scale=1]{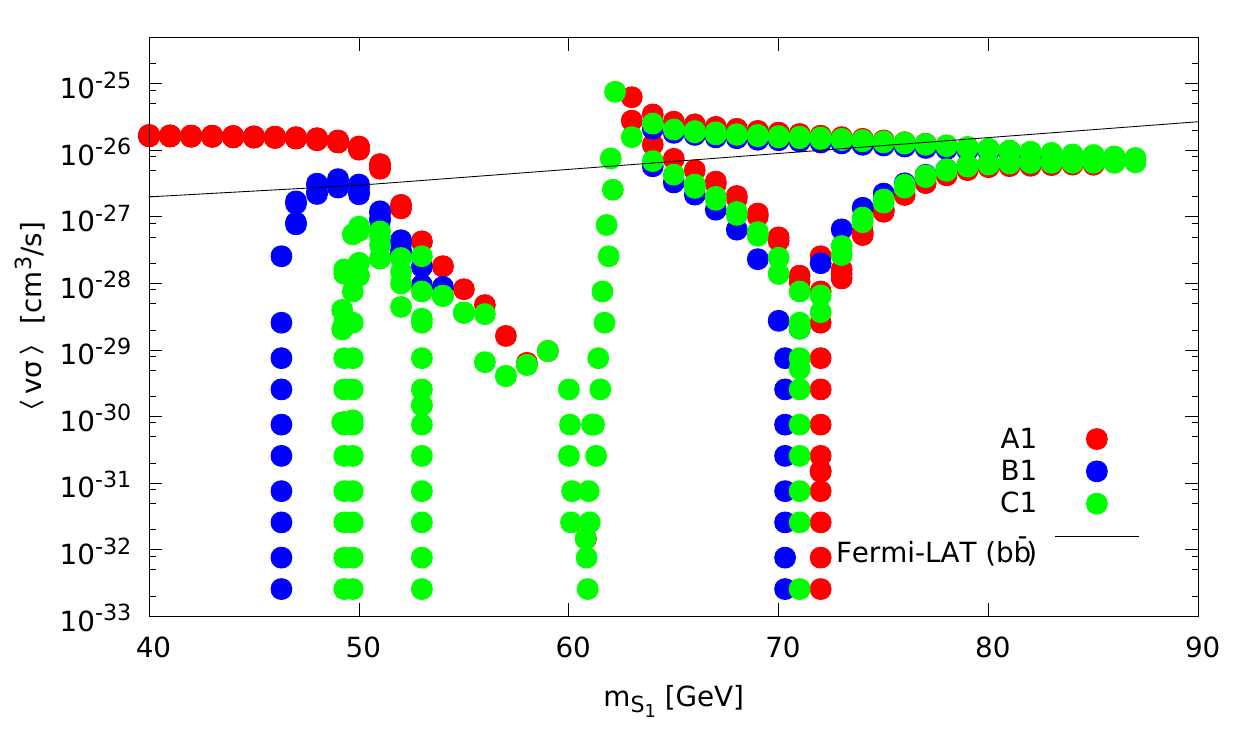}
\caption{Indirect detection limits for low and medium mass region.\label{indirect}}
\end{figure}

Recent indirect detection results from Fermi-LAT strongly constrain the DM candidate annihilating into $b\bar{b}$ pair \cite{Ackermann:2015zua}, and therefore are  crucial for the low DM mass region. The CP-conserving scalar Higgs-portal type of DM with proper relic density and $m_{S_1} \lesssim 53$ GeV is ruled out \cite{Duerr:2015aka}. The same limit applies to case A1, as the dominant annihilation channel is into $b\bar{b}$ pair (Figure \ref{indirect}). 

For cases B1 and C1 annihilation channels are different and good relic density is obtained for smaller values of Higgs-DM coupling. This weakens the annihilation into $b\bar{b}$, leading to most of the parameter space to lie within the allowed region.

For A1, B1 and C1 the resonance region for $m_{S_1} < m_h/2$ is in agreement with Fermi-LAT constraints.

Fermi-LAT results will also constrain the medium mass region, although in the less stringent way than in case of the standard Higgs-portal DM model. Region just above the Higgs-resonance can be excluded by the indirect detection results, as the main annihilation channel for DM candidate is annihilation into $b \bar{b}$ pair of the order of $10^{-26} {\rm cm}^3/$s. For heavier masses, i.e. $m_{S_1} \gtrsim 66 \GeV$ annihilation into gauge bosons starts to be of the same order as the $b\bar{b}$, and then quickly dominates over all other annihilation channels. The annihilation cross section gets smaller, of the order of $10^{-27} cm^3/$s. In Fig. \ref{indirect} one can see two branches, corresponding to two regions of good relic density from Fig. \ref{relic2}. The upper branch, which corresponds to the lower branch in Fig.\ref{relic2} (i.e. with larger values of $|g_{S_1 S_1 h}|$) is excluded by the indirect DM detection results. The lower branch, especially the region of masses which need $g_{S_1 S_1 h} \approx 0$ escapes this constraint.

For the heavy DM candidate constraints for the parameter space of the heavy DM candidate may come from the indirect detection experiments, and they provide a complementary way to constrain the region. Analysis performed in \cite{Queiroz:2015utg,Garcia-Cely:2015khw} shows that the H.E.S.S. experiment can already test the parameter space of the IDM, which in the heavy mass region is similar to the case H1 of I(2+1)HDM. Also, the upcoming Cherenkov Telescope Array will be able to probe a significant part of the high mass regime of the models like the IDM or the I(2+1)HDM, testing masses of DM candidate up to 800 GeV. 

\subsubsection{Interplay between direct and indirect detection experiments}

Direct and indirect detection experiments provide a complementary way to constrain the parameter space of the model, see Table \ref{detection_ex}. It is especially important for masses just above $m_h/2$, which escapes the possibility of direct detection, however, due to an enhancement from the Breit-Wigner resonance effect it is possible to exclude this region from the results of indirect detection experiments. 

\begin{table}[]
\centering
\caption{Exclusions from direct and indirect detection experiments.}
\label{detection_ex}
\begin{tabular}{|l|l|l|l|}
\cline{1-4}
 benchmark & $m_{S_1}$ & DD & ID \\ \cline{1-4}
 A1 &$m_{S_1} \lesssim 53 \GeV$ & $\times$ & $\times$ \\ \cline{1-4}
 B1 &$m_{S_1} \lesssim 53 \GeV$ & $\times$ & $\surd$ \\ \cline{1-4}
  C1 &$ 49 \lesssim m_{S_1} \lesssim 53 \GeV$ & $\surd$ & $\surd$ \\ \cline{1-4}
 A1,B1,C1 & $53 \GeV \lesssim m_{S_1} \lesssim  m_h/2$ & $\surd$ & $\surd$ \\ \cline{1-4}
 A1,B1,C1 & $m_h/2 \lesssim m_{S_1} \lesssim  64 \GeV$ & $\surd$ & $\times$ \\ \cline{1-4}
 A1,B1,C1 & $64 \GeV \lesssim m_{S_1} \lesssim 74 \GeV$ & $\surd$ & $\surd$ \\ \cline{1-4}
 A1,B1,C1 & $74 \GeV \lesssim m_{S_1} \lesssim m_Z$ & $\times$ & $\surd$ \\ \cline{1-4} 
\end{tabular}
\end{table}

\subsection{LHC limits}
\subsubsection{Higgs inert decays and Higgs total decay strength}

\begin{figure}[h!]
\centering
\includegraphics[scale=1]{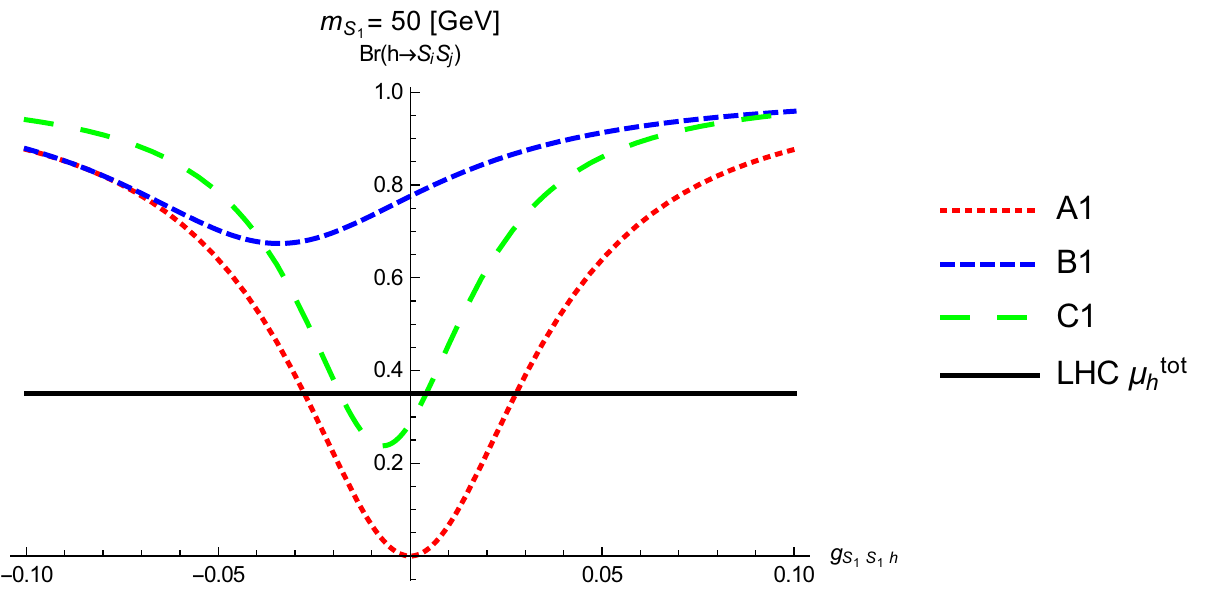}
\caption{Contribution to the Higgs invisible decays for $m_{S_1} = 50$ GeV. \label{invisible50}}
\end{figure}

Figure \ref{invisible50} presents the contribution to ${\rm BR}(h\to S_i S_j)$ for $m_{S_1} = 50$ GeV for cases A1, B1 and C1, following relation (\ref{inv_all}). Also, the limit from $\mu_{tot}$ is shown. In case A1 there is only one particle that contributes to the Higgs decay (the DM candidate $S_1$). For small values of $g_{S_1 S_1 h}$ the contribution to the total decay width of the Higg particle is small enough. There is also a small region fulfilling this constraint for case C1, but not for case B1. One would expect that case C1, where there are up to four light particles would have a bigger branching ratio. However, as seen in Figs.\ref{couplingsh1} and \ref{couplingsh2}, case C1 has actually smaller values of Higgs-inert couplings than case B1.

\begin{figure}[h]
\centering
\includegraphics[scale=1]{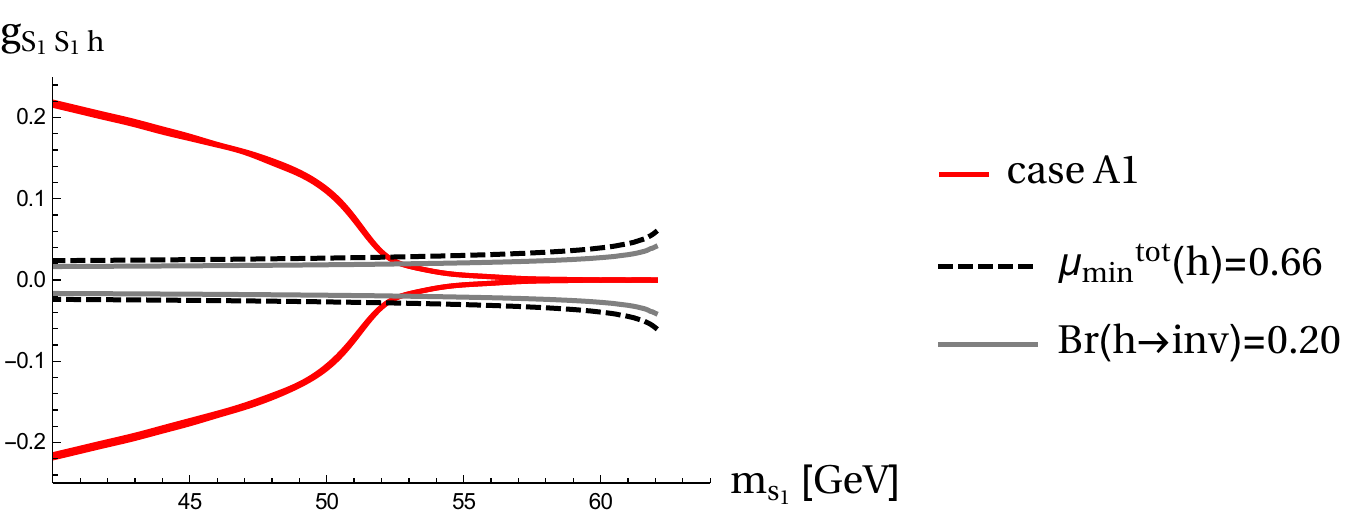}
\caption{Relic density constraints vs. Higgs invisible branching ratio and Higgs total signal strength bounds for scenario A1. \label{invA}}
\end{figure}

\begin{figure}[h!]
\centering
\includegraphics[scale=1]{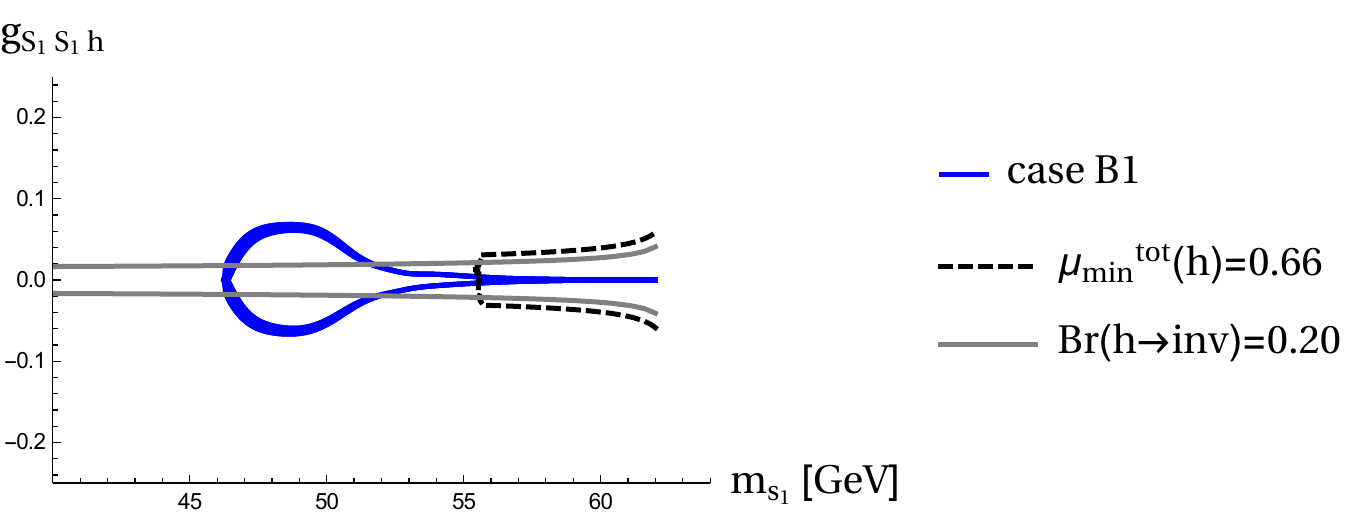}
\caption{Relic density constraints vs. Higgs invisible branching ratio and Higgs total signal strength bounds for scenario B1. \label{invB}}
\end{figure}

\begin{figure}[h!]
\centering
\includegraphics[scale=1]{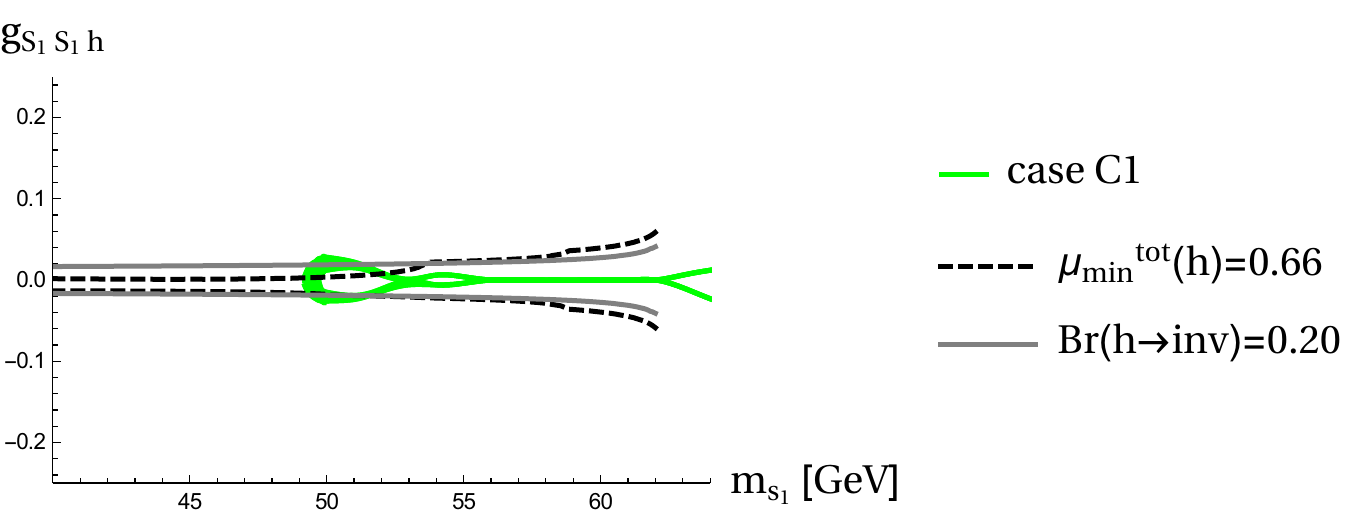}
\caption{Relic density constraints vs. Higgs invisible branching ratio and Higgs total signal strength bounds for scenario C1.  \label{invC}}
\end{figure}

Figs. \ref{invA}, \ref{invB}, \ref{invC} show constraints from the Higgs invisible branching ratio (${\rm BR}(h\to inv) = 0.2$) and $\mu_{tot}$ for scenarios A1, B1 and C1. The solid line corresponds to the limit for ${\rm BR}(h\to inv)$ following Eq. \ref{invs1}. Generally, $g_{S_1 S_1h}$ has to be small. This limit, applied to results from Fig. \ref{relic1}, constrains the masses of DM particle and benchmark points.

We want to stress that the LHC limits provide stronger constraints for some benchmark points in the low mass region than the dedicated DM detection experiments. It is especially important considering the astrophysical uncertainties that may influence interpretation of results provided by DM detection experiments. By using the LHC data we can test the model without relying on them.

\subsection{$h\to\gamma\gamma$ signal strength}

\begin{figure}[h!]
\centering
\includegraphics[scale=1]{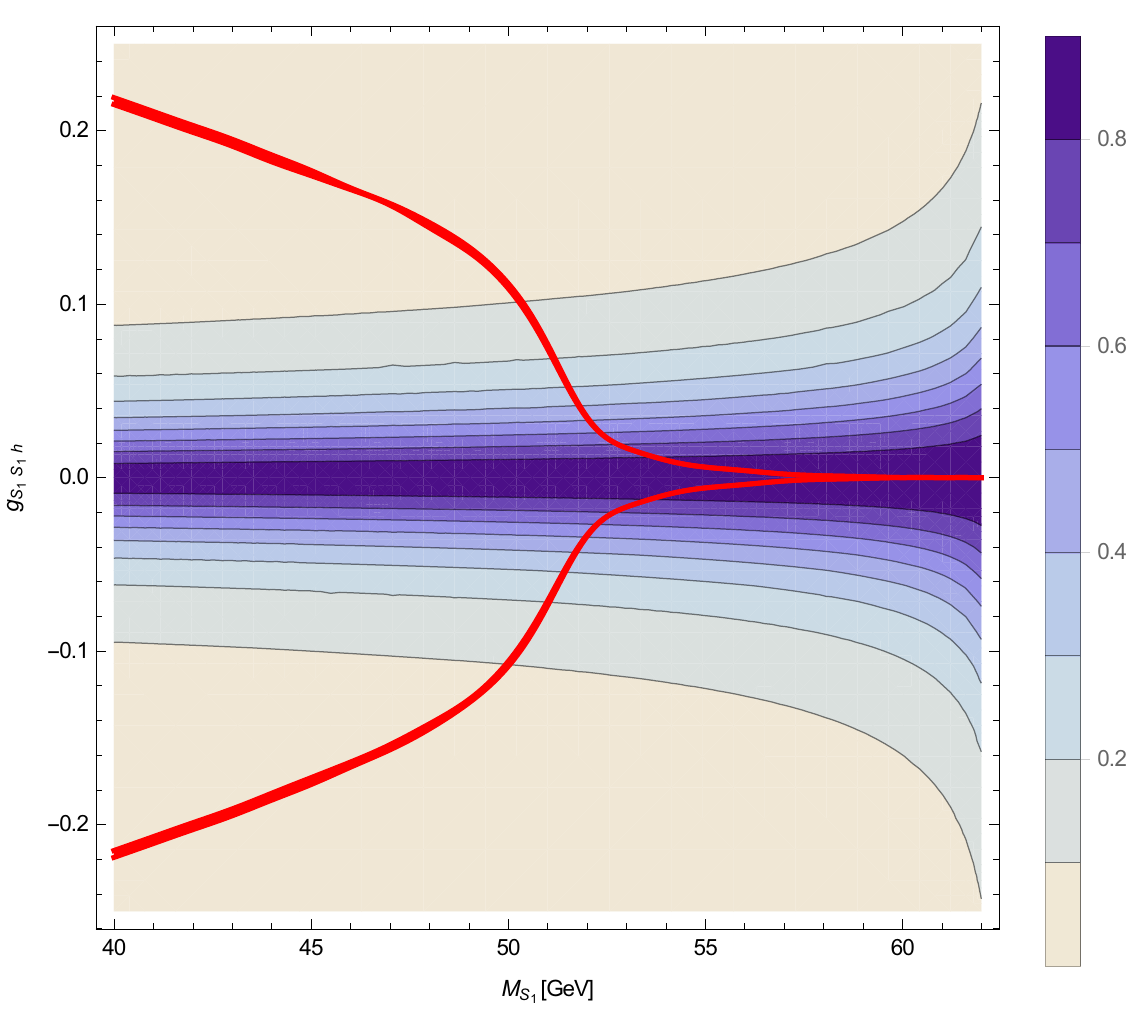}  \includegraphics[scale=1]{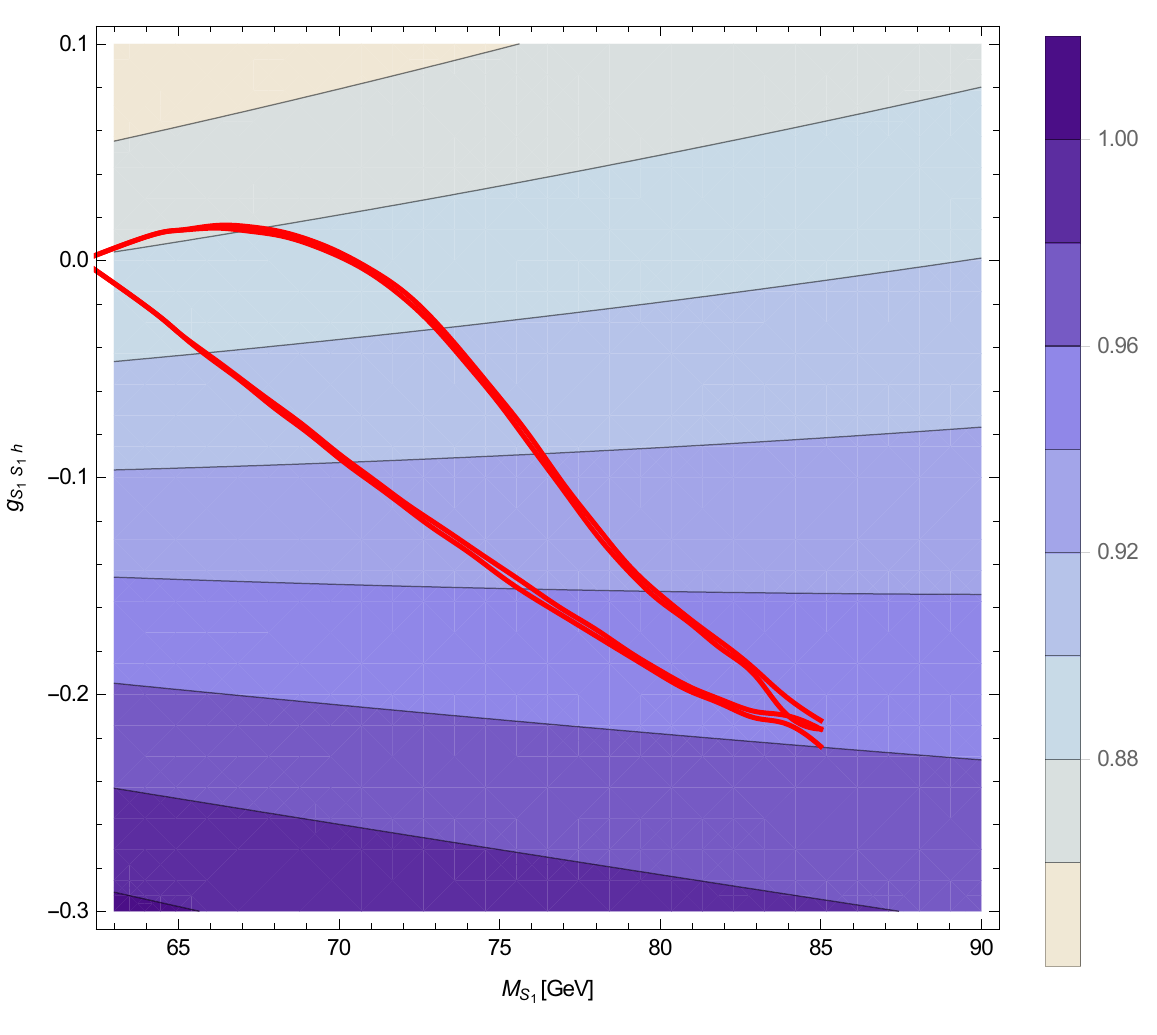}
\caption{$h\gamma\gamma$ signal strength with relic density limits for scenario A1. \label{hgg_a1_low}}
\end{figure}

\begin{figure}[h!]
\centering
\includegraphics[scale=1]{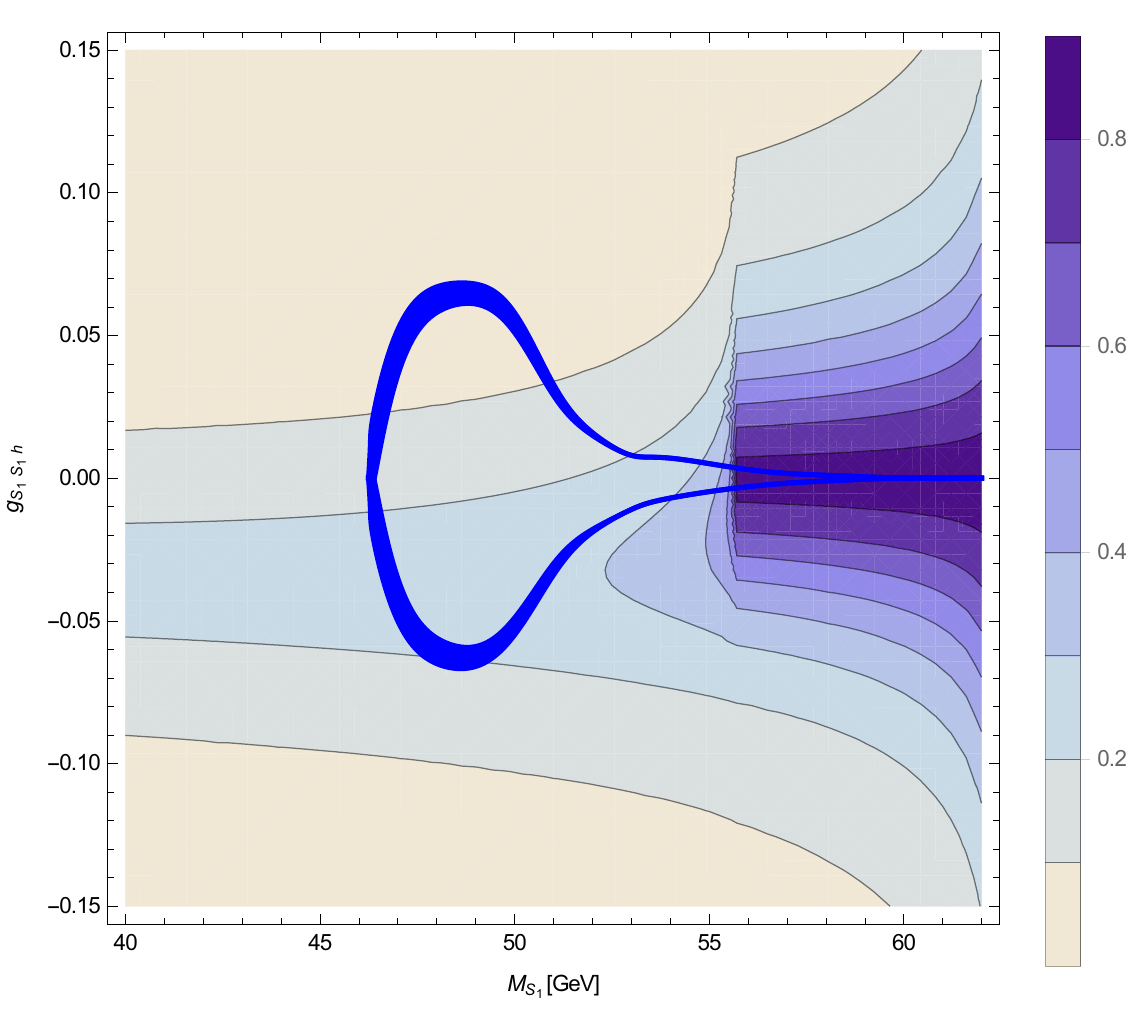} 
\includegraphics[scale=1]{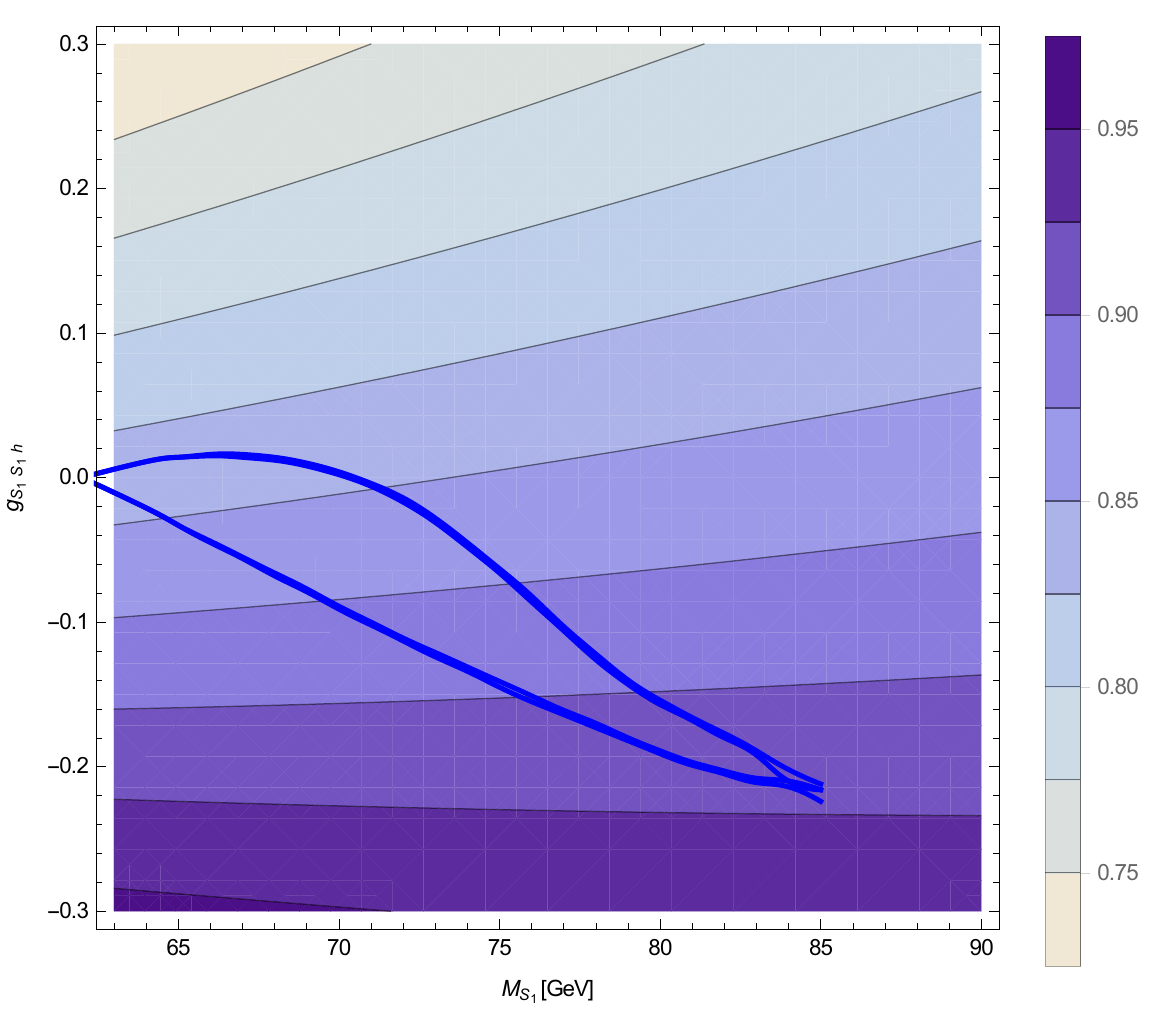}
\caption{$h\gamma\gamma$ signal strength  with relic density limits for scenario  B1. \label{hgg_b1_low}}
\end{figure}

\begin{figure}[h!]
\centering
\includegraphics[scale=1]{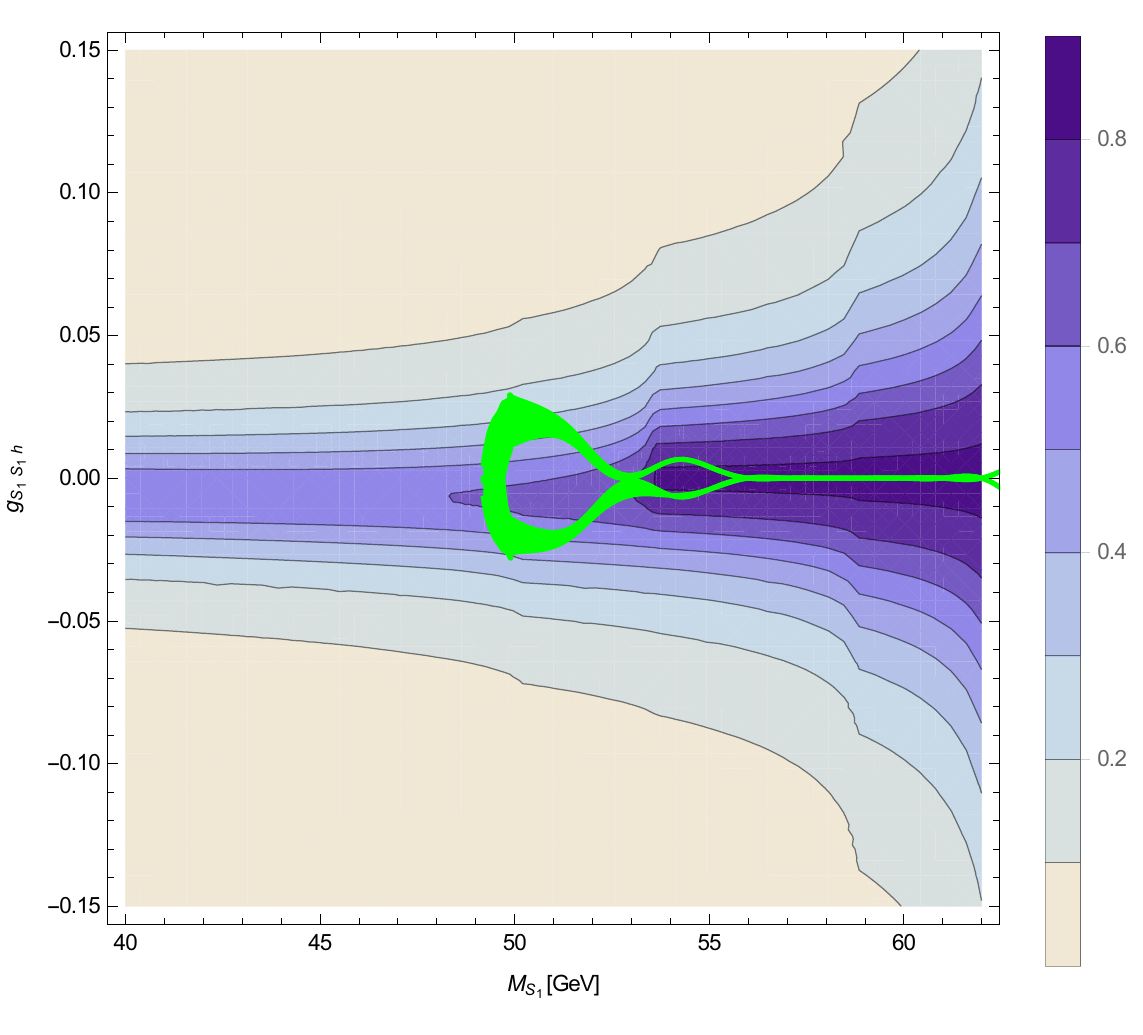} 
\includegraphics[scale=1]{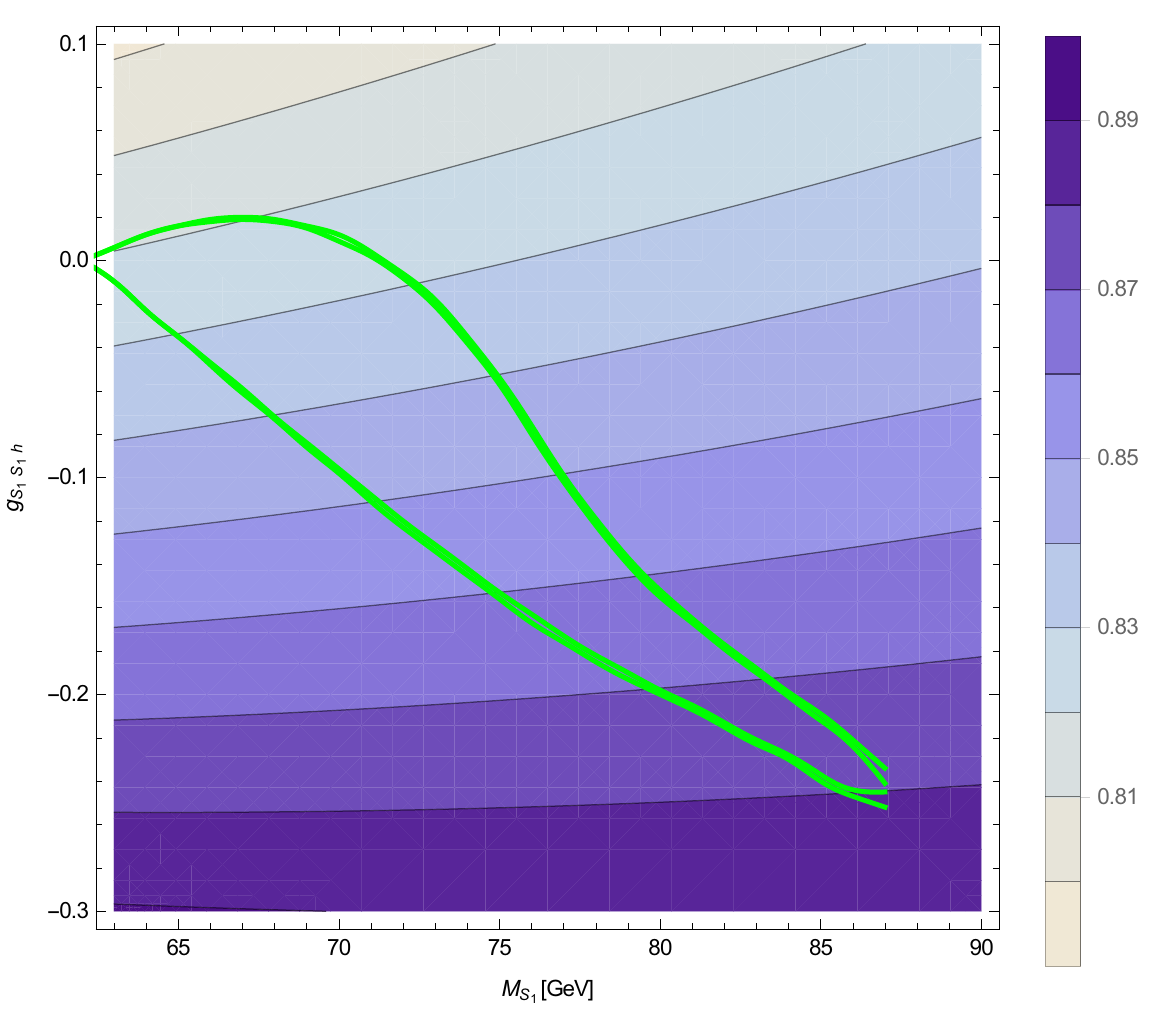}
\caption{$h\gamma\gamma$ signal strength  with relic density limits for scenario  C1. \label{hgg_c1_low}}
\end{figure}

\begin{figure}[h!]
\centering
\includegraphics[scale=1]{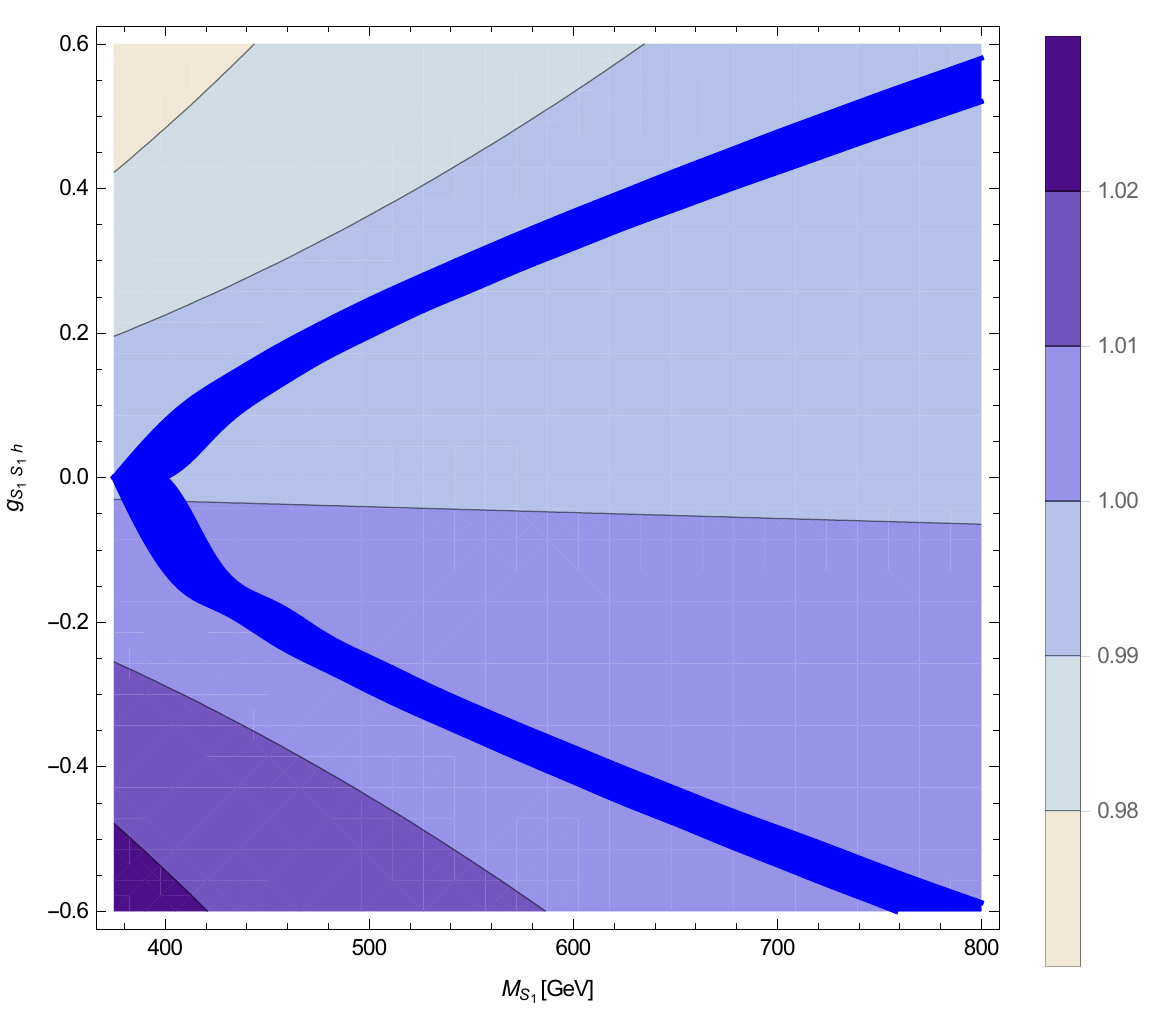} \includegraphics[scale=1]{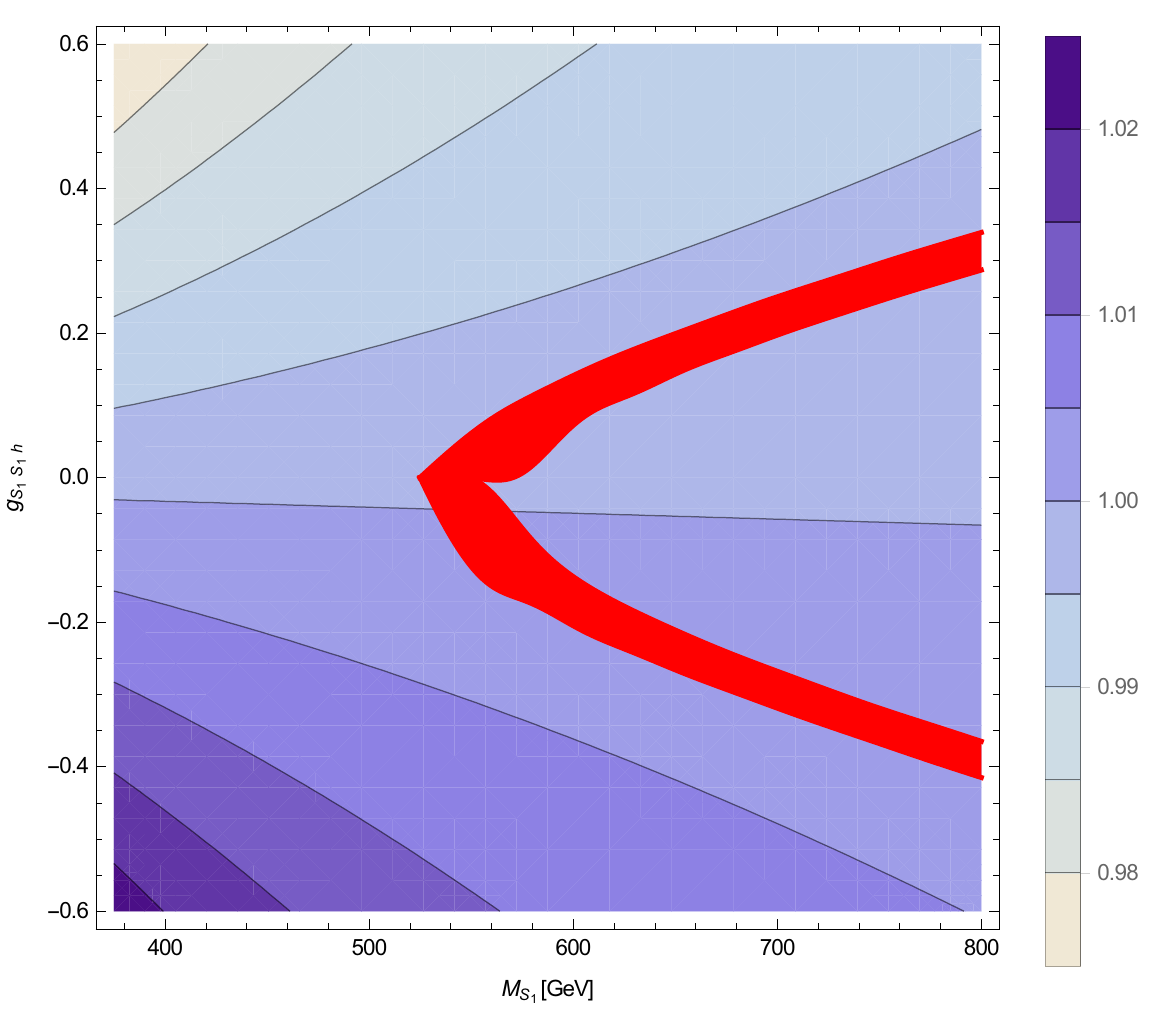}
\caption{$h\gamma\gamma$ signal strength  with relic density limits for scenario  G1 (top) and H1 (bottom). \label{hgg_heavy}}
\end{figure}

\begin{figure}[h!]
\centering
\includegraphics[scale=1]{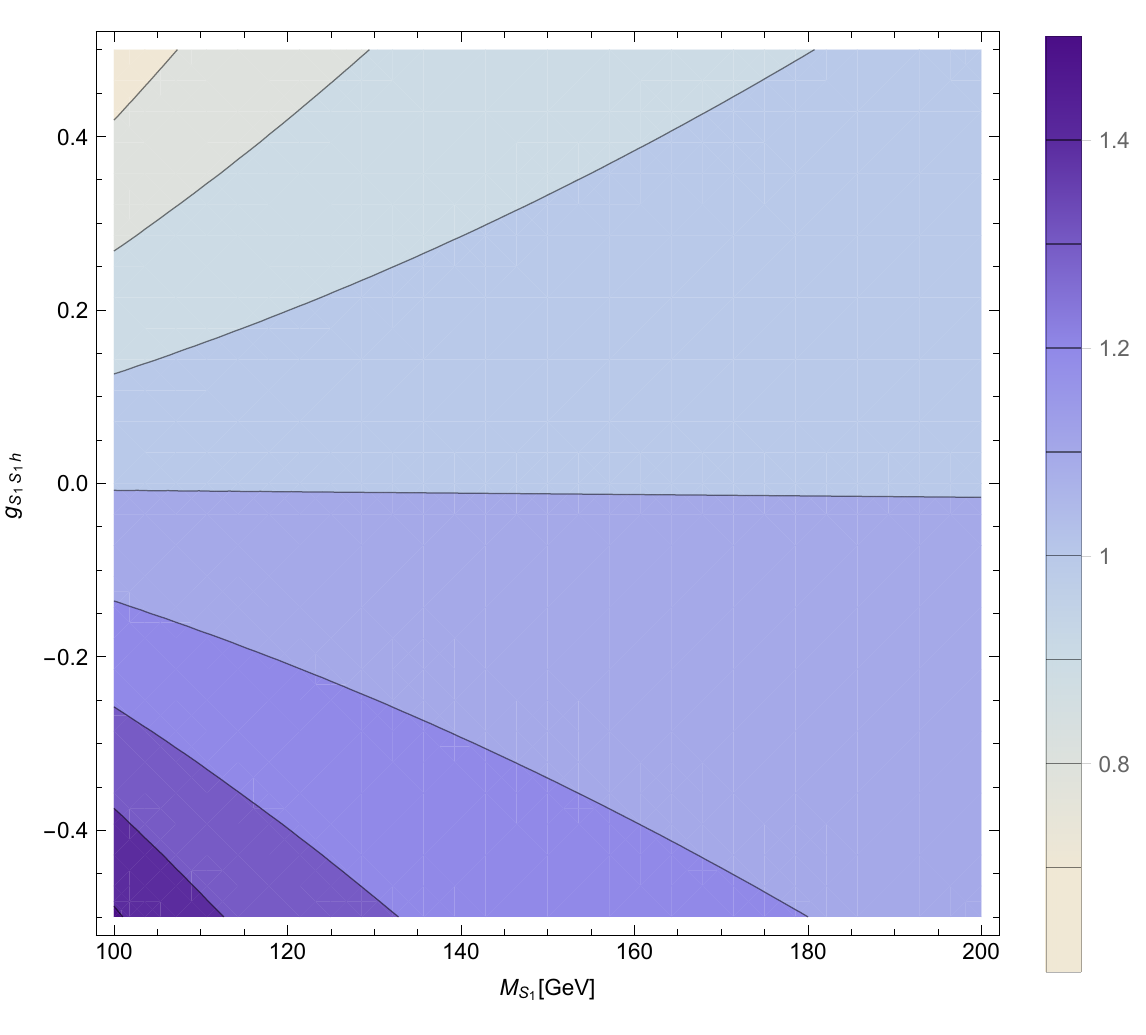} \includegraphics[scale=1]{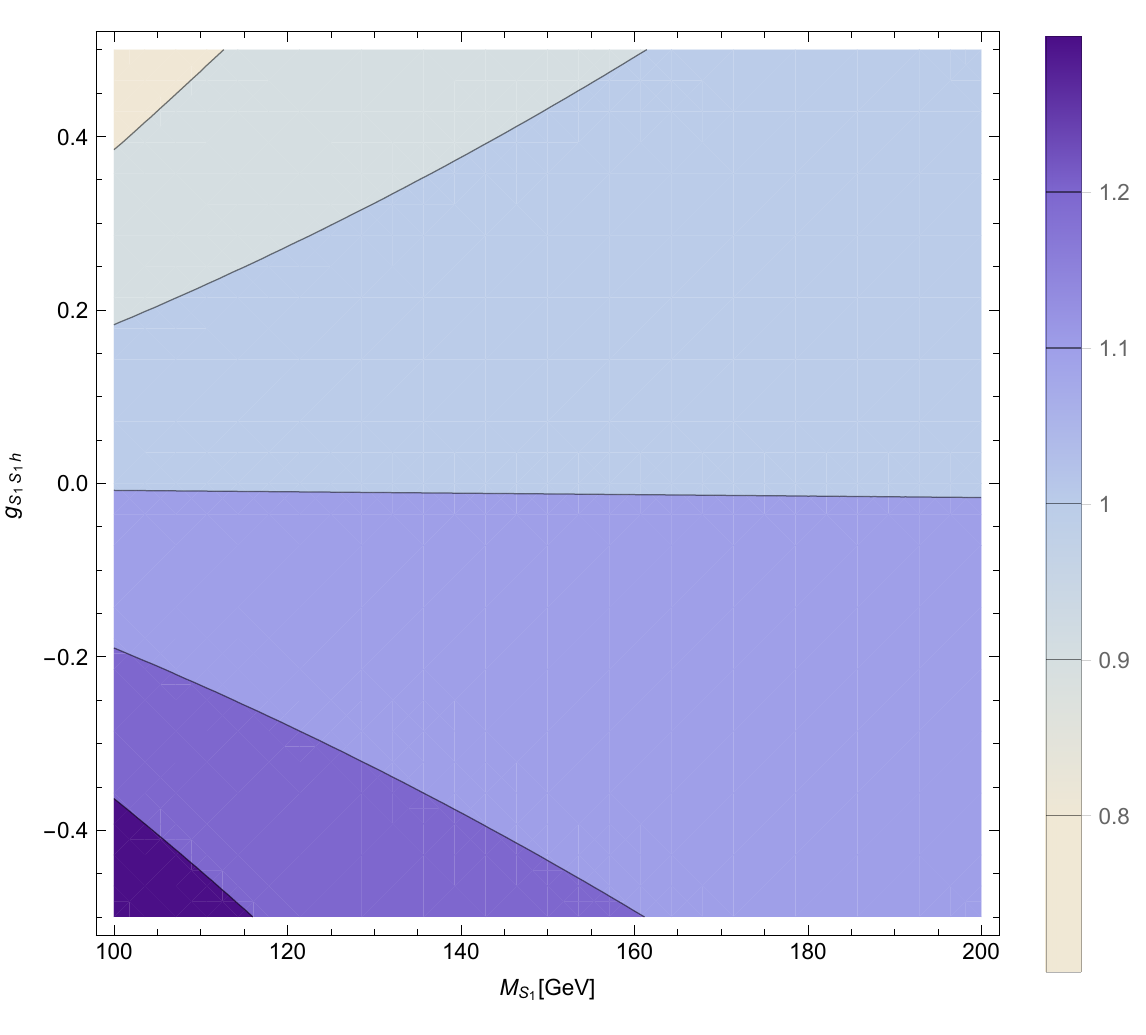}
\caption{$h\gamma\gamma$ signal strength for G1 (top) and H1 (bottom) in the medium mass region (relic density in this region is below the Planck limit). \label{hgg_medium}}
\end{figure}

Strong constraints come from $h\to\gamma\gamma$ signal strength data.
\begin{enumerate}
\item In the low mass region Higgs to $\gamma \gamma$ signal strength is heavily influenced by the presence of light neutral particles. The contribution to the total decay width of the Higgs is so strong, that it is not possible to compensate this change by an increase in the partial decay width $h \to \gamma \gamma$. It it seen in Fig. \ref{hgg_a1_low}, \ref{hgg_b1_low} and \ref{hgg_c1_low}, where the maximum value of $\mu_{\gamma \gamma}$ is around 0.9 for small values of Higgs-DM couplings. It is also clear that this cosntraint, related to limits for Higgs total decay width from Fig.\ref{invA}, \ref{invB} and \ref{invC}, is limiting the parameter space very strongly. Exclusion limits for case B1 are much stronger than these obtained from direct or indirect detection experiments. 

\item In the medium mass region the additional decay channels are closed, leading to a possibility of enhancement in the $\gamma \gamma$ channel. However, our study shows that for values of couplings that give good relic density, the $\mu_{\gamma \gamma}$ is still below the SM value, although it is closer to it than in the low DM mass region. Values are bigger for case C1, where there are two charged scalar particles with similar masses. As discussed before, contribution from the heavier scalar is smaller than from $S_1^\pm$.

\item Fig. \ref{hgg_medium} present the calculation of $\mu_{\gamma \gamma}$ for benchmarks G1 and H1, but for the DM mass between 100 and 200 GeV. With this choice of parameter the relic density is too small and it is not a viable region of parameter space (unless one accepts the possibility of having a subdominant DM candidate, which we are not discussing here). For this choice of parameters two charged scalars are very close in mass and they are relatively light. This means that their contribution to the $h\gamma \gamma$ loop is large, and indeed one can see the significant enhancement in this channel.
\item Fig. \ref{hgg_heavy} shows the only region where it is possible to have a good relic density, and $\mu_{\gamma \gamma}$ equal to at least the SM value. The enhancement is there, although it is minimal. It is related to having much heavier charged scalars than $m_h$.
\item If the measured value of $\mu_{\gamma \gamma} > 1$ then only heavy DM mass region will survive, unless we accept the subdominant DM candidate or find a region between about 100-200 GeV with good relic density. Within the experimental error we can find solutions in all studied regions. 
\item We would like to stress that there is a tension with direct and indirect detection limits in the medium mass region. To have a larger value of $\mu_{\gamma \gamma}$ we need to have a negative coupling with a relative large absolute value. This means that we need to be on the lower branch in Fig. \ref{relic2}, which corresponds to the upper branch in Fig. \ref{direct1} and \ref{indirect}. 

\end{enumerate}

\section{Conclusion and outlook}
\label{conclusion}

In this paper we have studied an extension of the Standard Model (SM) in which two copies of the SM-Higgs doublet which do not acquire a Vacuum Expectation Value (VEV), and hence are \textit{inert}, are added to the scalar sector. In other words, this is a 3HDM with two inert and one active scalar doublet,
denoted as the I(2+1)HDM.
We have allowed for CP-violation in the \textit{inert} sector, where the lightest \textit{inert} state
is protected from decaying to SM particles through the conservation of a $Z_2$ symmetry. 
The lightest neutral particle from the \textit{inert} sector, which has a mixed CP-charge due to CP-violation, is hence a DM  candidate in the model. 

After giving the scalar potential, we have calculated the mass spectrum in
the ``dark democracy'' limit, in which the two inert doublets are treated on an equal footing,
in order to simplify the parameter space of the model. For instance, in this limit, CP violation
in the inert sector is controlled by only a single angle $\theta_2 + \theta_{12}$.
After considering various theoretical and experimental constraints on the parameter space of the model,
using recent results from the LHC and DM direct and indirect detection experiments, we then focussed on five representative benchmark scenarios relevant for DM studies.

We then discussed the new regions of DM relic density opened up by CP-violation,
for the chosen benchmark scenarios, defining three benchmark points A1, B1, C1
in the low and medium DM mass region (below the $Z$ mass)
and two points G1, H1 in the high DM mass region (above 400 GeV),
comparing our results to the IDM in all cases.
We find that with the introduction of CP violation, the strength of the couplings which were fixed in the CP conserving limit, become unconstrained. 
Regarding relic density studies, with CP violation,
scenarios B and C populate the complete region of Higgs-DM coupling between zero and what was accessible in the CP conserving limit. We show that the direct and indirect detection experiments which excluded most of the parameter space in the low mass region in the CP conserving limit, leave scenario C uncut due to the very small Higgs-DM coupling in such scenarios.

The most constraining bounds come from the LHC data. This is where the CP-violating scenarios
differ most significantly from the CP-conserving case, since scenarios C allow for the Higgs-DM coupling to be close to zero passing all LHC bounds.
In the medium mass region all three scenarios A, B and C have the same relic density behaviour as the CP conserving limit. The data from h$\gamma\gamma$ signal strength shows a tendency for heavier DM mass in this region. In the heavy mass region, the CP violating scenarios behave the same as the CP conserving limit. According to the data from h$\gamma\gamma$ signal strength this region is preferred for the DM mass.
The LHC signatures of this model will be explored further in a future publication.

\section*{Acknowledgement}
SFK acknowledges support from the STFC Consolidated grant ST/L000296/1 and the
European Union Horizon 2020 research and innovation programme under the Marie 
Sklodowska-Curie grant agreements InvisiblesPlus RISE No. 690575 and 
Elusives ITN No. 674896.
SM is financed in part through the NExT Institute and from the STFC Consolidated ST/ J000396/1. He also acknowledges the H2020-MSCA-RICE-2014 grant no. 645722 (NonMinimalHiggs).
VK's research is financially supported by the Academy of Finland project ``The Higgs Boson and the Cosmos'' and  project 267842.
DS is partially supported by the HARMONIA project under contract UMO-2015/18/M/ST2/00518 (2016-2019).
JHS, DR and AC are supported by CONACYT (M\'exico), VIEP-BUAP and 
PRODEP-SEP (M\'exico) under the grant: ``Red Tem\'atica: F\'{\i}sica del Higgs y del Sabor".

\end{document}